\documentclass{dune}
\pdfoutput=1

\usepackage[utf8]{inputenc}
\usepackage{mathtools}
\usepackage{tikz}
\usepackage[compat=1.1.0]{tikz-feynman}
\usetikzlibrary{shapes,arrows,positioning,automata,backgrounds,calc,er,patterns}

\usepackage[colorlinks = true,
            linkcolor = black,
            urlcolor  = black,
            citecolor = black,
            anchorcolor = black]{hyperref}
            
\usepackage{array, booktabs, color, eurosym, graphicx, longtable, pbox, pdfpages, url, verbatim}
\usepackage{wrapfig}
\usepackage[numbers,sort&compress]{natbib}
\usepackage{abstract}

\newcommand\bea{\begin{eqnarray}}
\newcommand\eea{\end{eqnarray}}

\colorlet{Changes@Color}{violet} 

\newbox\absbox
\renewenvironment{abstract}{\global\setbox\absbox=\vbox\bgroup
  \hsize=\textwidth%
  \noindent\unskip\textbf{Executive Summary}
 \par\medskip\noindent\unskip\ignorespaces}
 {\egroup}

\begin{document}

\title{\scshape\Large Cosmogenic Dark Matter and Exotic Particle Searches in Neutrino Experiments \\
{\normalsize NF03 Contributed White Paper to Snowmass 2021}
\vskip -10pt}

\begin{abstract}
In the past years, various new physics models predicting non-conventional cosmogenic signals have been proposed partly because no conclusive dark-matter signals have been observed in conventional dark-matter detection experiments and partly because a growing number of experiments including neutrino experiments feature kiloton-scale detectors which can be sensitive to such signals. 
\end{abstract}

%
%

\renewcommand\Authfont{\scshape\small}
\renewcommand\Affilfont{\itshape\footnotesize}

\author[7]{J.~Berger}
\author[8]{D.~Brailsford}
\author[1]{K.~Choi}
\author[5]{J.\ I.\ Crespo-Anad\'on}
\author[3]{Y.~Cui}
\author[12]{A.~Das}
\author[4]{J.A.~Dror}
\author[10]{Alec~Habig}
\author[11]{Y.~Itow}
\author[2]{E.~Kearns}
\author[13]{D.~Kim\thanks{Lead Editors}}
\author[6]{J.-C.~Park}
\author[12]{G.~Petrillo}
\author[15,17]{C.~Rott}
\author[9]{M.~Sen}
\author[14]{V.~Takhistov}
\author[12]{Y.-T.~Tsai\thanks{Lead Editors}}
\author[16] {J.~Yu}

\affil[1]{Center for Underground Physics, Institute for Basic Science (IBS), Daejeon 34047, Republic of Korea}
\affil[2]{Department of Physics, Boston University, Boston, MA 02215, USA}
\affil[3]{Department of Physics and Astronomy, University of California, Riverside, CA 92521, USA}
\affil[4]{Department of Physics, University of California Santa Cruz, Santa Cruz, CA 95064, USA and Santa Cruz Institute for Particle Physics, Santa Cruz, CA 95064, USA}
\affil[5]{CIEMAT, Centro de Investigaciones Energéticas, Medioambientales y Tecnológicas, E-28040 Madrid, Spain}
\affil[6]{Department of Physics, Chungnam National University, Daejeon 34134, Republic of Korea}
\affil[7]{Department of Physics, Colorado State University, Fort Collins, Colorado 80523, USA}
\affil[8]{Physics Department, Lancaster University, Bailrigg, Lancaster LA1 4YB, United Kingdom}
\affil[9]{Max-Planck-Institut f{\"u}r Kernphysik, Saupfercheckweg 1, 69117 Heidelberg, Germany}
\affil[10]{Department of Physics and Astronomy, University of Minnesota Duluth, Duluth, MN 55812, USA}
\affil[11]{Solar-Terrestrial Environment Laboratory, Nagoya University, Nagoya, Japan}
\affil[12]{SLAC National Accelerator Laboratory, Menlo Park, CA 94025, USA}
\affil[13]{Mitchell Institute for Fundamental Physics and Astronomy, Texas A\&M University, College Station, TX 77843, USA}
\affil[14]{Kavli Institute for the Physics and Mathematics of the Universe (WPI), UTIAS \\The University of Tokyo, Kashiwa, Chiba 277-8583, Japan}
\affil[15]{Department of Physics and Astronomy, University of Utah, Salt Lake City, UT 84112, USA}
\affil[16]{Department of Physics, University of Texas, Arlington, TX 76019, USA}
\affil[17]{Department of Physics, Sungkyunkwan University, Suwon 16419, Korea}

\date{}

\maketitle

\renewcommand{\familydefault}{\sfdefault}
\renewcommand{\thepage}{\roman{page}}
\setcounter{page}{0}

\pagestyle{plain} 
\clearpage
\textsf{\tableofcontents}


\renewcommand{\thepage}{\arabic{page}}
\setcounter{page}{1}

\pagestyle{fancy}

\fancyhead{}
\fancyhead[RO]{\textsf{\footnotesize \thepage}}
\fancyhead[LO]{\textsf{\footnotesize \rightmark}}

\fancyfoot{}
\fancyfoot[RO]{\textsf{\footnotesize Snowmass 2021}}
\fancyfoot[LO]{\textsf{\footnotesize Cosmogenic Dark Matter and Exotic Particle Searches in Neutrino Experiments}}
\fancypagestyle{plain}{}

\renewcommand{\headrule}{\vspace{-4mm}\color[gray]{0.5}{\rule{\headwidth}{0.5pt}}}

\newcommand{\dk}[1]{\textcolor{red}{#1}}
\newcommand{\dkc}[1]{\textcolor{red}{[#1 -- DK]}}



\section*{Executive Summary}
\label{sec:summary}

The signals from outer space and their detection have been playing an important role in particle physics, especially in discoveries of and searches for physics beyond the Standard Model (BSM); beyond the evidence of dark matter (DM), for example, the neutrinos produced from the dark matter annihilation is important for the indirect DM searches. 
Moreover, a wide range of new, well-motivated physics models and dark-sector scenarios have been proposed in the last decade, predicting cosmogenic signals complementary to those in the conventional direct detection of particle-like dark matter.
Most notably, various mechanisms to produce (semi-)relativistic DM particles in the present universe (e.g. boosted dark matter) have been put forward, while being consistent with current observational and experimental constraints on DM. 
The resulting signals often have less intense and more energetic fluxes, to which underground, kiloton-scale neutrino detectors can be readily sensitive.
In addition, the scattering of slow-moving DM can give rise to a sizable energy deposit if the underlying dark-sector model allows for a large mass difference between the initial and final state particles,
and the neutrino experiments with large volume detectors are well suited for exploring these opportunities. 

Detectors based on different technologies are complementary on probing diverse models and scenarios.
For example, water Cherenkov detectors normally have large mass, nanosecond-level time resolution, and MeV-level detection thresholds for electrons, leading to the most stringent limits to date on boosted dark matter originating from the Galactic Center and the Sun~\cite{Agashe:2014yua, Berger:2014sqa, Kong:2014mia, Super-Kamiokande:2017dch} and on ``dark cosmic rays'' of DM accelerated in astrophysical sources.
While long-string water Cherenkov detectors are uniquely suitable for TeV-scale signals, liquid-argon time-projection chambers (LArTPCs) may have advantage on search for hadronic boosted DM interactions, owing to their moderately large nuclei and the capability of detecting protons with the kinetic energy down to a few tens of MeV.
As various neutrino experiments are currently collecting data or will be operational in the future, a vast swath of parameter space will soon be explored.

Spanning over a wide energy range, the cosmogenic BSM searches broaden the physics cases at neutrino detectors and enhance the research and development of experimental techniques and analysis strategies.
Neutrinos from natural sources, such as solar, atmospheric, and astrophyiscal neutrinos, generically contribute as the main background, and understanding of such neutrino fluxes and interaction cross sections are crucial.
Searches for MeV-scale signals also encounter the background sources from radiological materials, which make recording such signals in real time challenging.
Innovative development on detectors, triggering systems, reconstruction algorithms, etc. will be helpful to comprehensively collect and analyze physics activities of interest.

This White Paper is devoted to discussing the scientific importance of the cosmogenic dark matter and exotic particle searches, not only overviewing the recent efforts in both the theory and the experiment communities but also providing future perspectives and directions on this research branch. 
A landscape of technologies used in neutrino detectors and their complementarity is discussed, and the current and developing analysis strategies are outlined.

\newpage

\section{Introduction}
\label{sec:intro}

While the Standard Model (SM) has successfully described the interactions of elementary particles and force carriers, it leaves compelling questions unanswered. For instance, it fails to explain the number and mass differences of elementary particles, and the observed matter-antimatter asymmetry in the universe.  In addition, recent experiments and observations have revealed a few phenomena beyond the SM (BSM), such as dark matter (DM) and evidence of neutrino masses.
Over the past decades, a lot of theoretical and experimental efforts have been made to address these questions, utilizing both the artificial and natural sources.
In particular, observation of outer space signals has been an important probe in physics for hundred years, 
including the progresses on DM (in)direct searches and high-energy cosmic-ray measurements.
The cosmogenic sources, along with advancing technologies, are expected to keep offering unprecedented opportunities on BSM physics.

For the last decade, a variety of scenarios and models giving rise to cosmogenic DM and exotic particle signals have been proposed and developed, and the associated phenomenology has been extensively investigated.
In particular, null signal observation at DM detection experiments mostly aiming at weakly-interacting-massive-particle (WIMP) dark-matter candidates greatly motivates alternative possibilities that often involve non-minimal dark-sector scenarios beyond WIMP and result in non-conventional signals coming from the present universe.
While being consistent with existing bounds and constraints, the expected signal flux associated with these models is small, often requiring advanced detectors, such as those with a large mass or high resolution.

The development and results on neutrino experiments, detecting neutrinos produced by both natural and artificial sources, have been recently skyrocketing, to answer the main questions in neutrino physics, such as CP-violation in neutrino sector, neutrino mass and its ordering, the number of neutrino flavors, etc.
Owing to the weakly interacting nature of neutrinos and the finite intensity of neutrino fluxes, long-baseline neutrino experiments typically aim for large detector mass (e.g. multi-kiloton scale or larger).
The far detectors of these experiments are often placed deep underground to minimize the background activities with higher interaction rates. 
The detectors feature high resolution in measuring physical quantities, such as energy, timing, interaction position.
In particular, the recent development on detector technology provides better tracking capabilities,
significantly improving measurements of multi-particle final states. 
As a consequence, the neutrino detectors are greatly apt for the detection of cosmogenic BSM signals, including complicated signatures arising in non-minimal dark-sector scenarios.
A number of phenomenological studies have demonstrated promising BSM sensitivities at neutrino experiments,
while several neutrino experiments have reported dedicated searches for unconventional cosmogenic DM, expanding the limits in the related models~\cite{Super-Kamiokande:2017dch, COSINE-100:2018ged, PandaX-II:2021kai, CDEX:2022fig}.

The objective in this White Paper is to collect both theoretical and experimental efforts and to lay out future perspectives about searches for cosmogenic BSM signals, including dark matter and generic exotic particles, in neutrino experiments.
While a wide range of physics topics in this research area have been actively emerging and motivating the experimental community,
we expect that this White Paper to increase the awareness of the physics community for the ongoing efforts and future directions, and thus invigorate more research and communications between the theoretical/phenomenological and the experimental communities.

This White Paper is organized as follows. 
We begin in Section~\ref{Sec:models} with a brief overview of physics models predicting cosmogenic BSM signals, followed by a survey of individual scenarios and related phenomenology in Section~\ref{sec:modeloverview}.
Categorized by the detector technology, 
relevant neutrino detectors which would be sensitive to the cosmogenic signals are outlined in Section~\ref{sec:exp}.
We then lay out analysis strategies for the cosmogenic signal searches in Section~\ref{sec:strategy}, including signal and background simulations, triggering, and event reconstruction strategies.
Section~\ref{sec:complement} provides complementary aspects of cosmogenic new physics searches at neutrino experiments to other experiments, e.g. accelerator-based experiments. 
Finally, our conclusions and outlook appear in Section~\ref{sec:conclusion}.

\section{Theory Landscape of Cosmogenic Signals \label{Sec:models}  }
The discovery of DM in the universe is one of the great triumphs in particle physics, which is supported by various evidence in a wide range of observational scales, from the galactic scale to the cosmological scale. 
The SM does not explain DM and its related phenomena, and hence needs to be extended to include it. 
All of the evidence is based on its gravitational interactions, and the next task is to understand its particle nature through its hypothetical non-gravitational interactions with SM particles. 
While its existence is clearly confirmed, the mass scale of DM remains undecided, and, in principle, a wide range of mass values are allowed.
This does not necessarily imply that all mass values are equally motivated, but DM candidates of particular mass ranges have received more attention than the others. 
WIMP in the GeV-to-TeV mass range is one of such well-motivated DM candidates, as it is thermally produced and very predictive. 
In light of this situation, an enormous amount of experimental endeavor including DM (in)direct detection experiments and accelerator-based searches has been devoted to the detection of WIMP signals for the last few decades again via their non-gravitational interactions. 

When it comes to the indirect search, depending on the DM model details, relic DM including WIMP can annihilate or decay to neutrinos, and terrestrial neutrino detectors or telescopes are expected to observe such neutrino signals over the neutrinos originating from the known astrophysical sources. 
Due to the weakly interacting nature of neutrinos, their flux is least affected by the galactic medium and thus the signals can be readily associated with the source points. 
Currently, a number of related neutrino experiments have searched for or are planning to detect the DM-induced neutrinos, being now considered as traditional cosmogenic BSM signal searches.  

Beyond the neutrino signals emerging from the annihilation or decay of DM, a variety of models predicting cosmogenic, non-conventional DM and/or exotic particle signals have been suggested for the last decade, largely motivated by the null observation of WIMP signals.
Among them, scenarios of cosmogenic boosted dark matter (BDM) and related phenomenology have been most extensively investigated, including semi-annihilation, models of two-component DM, and cosmic-ray/neutrino-induced BDM.  
The common underlying idea is that a certain fraction of DM (or DM components) can be relativistic or fast-moving in the present universe, whereas the usual (cold) halo DM is somehow secluded from the SM sector or less sensitive to the existing experiments. 
Therefore, a promising strategy lies in searching for ``relativistic'' signatures typically involving energy deposits larger than the usual nuclear recoils induced by conventional non-relativistic DM.  

Another class of DM scenarios predicting large energy deposits is the so-called ``explosive'' DM.
Examples include nuclear destruction, self destruction and fermionic absorption. Unlike the BDM scenarios, the incoming DM species is still non-relativistic or slowly moving.
However, it is not absolutely stable and the relevant scattering processes are exothermic. As a result, the kinematics of the processes is governed by mass energy, not kinetic energy, naturally giving rise to large energy deposits inside the detector. 
Interestingly enough, explosive DM often carries a non-zero baryon number or lepton number or both, which may address the matter-antimatter asymmetry. 

In addition, energetic dark-sector particles can emerge in the atmosphere and astrophysical environments.
For example, millicharged dark-sector particles can be accelerated by astrophysical sources, e.g. supernova remnant, and energetic cosmic rays impinging on the atmosphere can copiously produce relativistic BSM particles.
These scenarios also predict energetic cosmogenic signals even beyond the PeV scale.    

The following section is devoted to a survey of relevant physics models and dark-sector scenarios: DM-induced neutrino signals in Section~\ref{sec:DMnus}, various mechanisms of producing BDM in the present universe in Section~\ref{sec:bdm}, exploding slow-moving DM in Section~\ref{sec:explosive}, and other cosmogenic signals from astrophysical sources and atmosphere in Section~\ref{sec:others}.

\section{Models and Phenomenology of Cosmogenic Signals \label{sec:modeloverview}}

\subsection{Dark-Matter-Induced Neutrinos}
\label{sec:DMnus}

\begin{figure*}
\begin{center}
\includegraphics[width=0.48\textwidth]{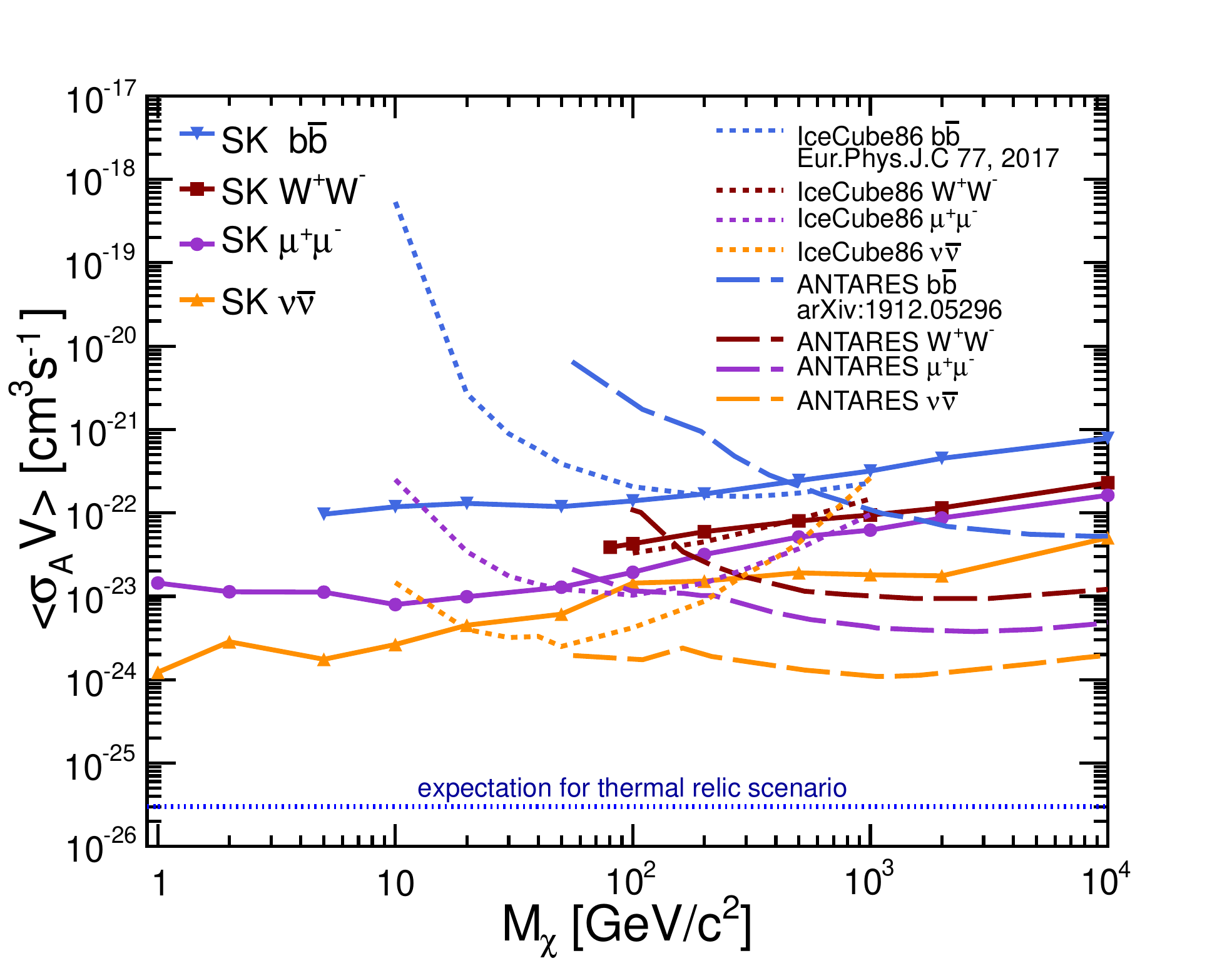} 
\includegraphics[width=0.48\textwidth]{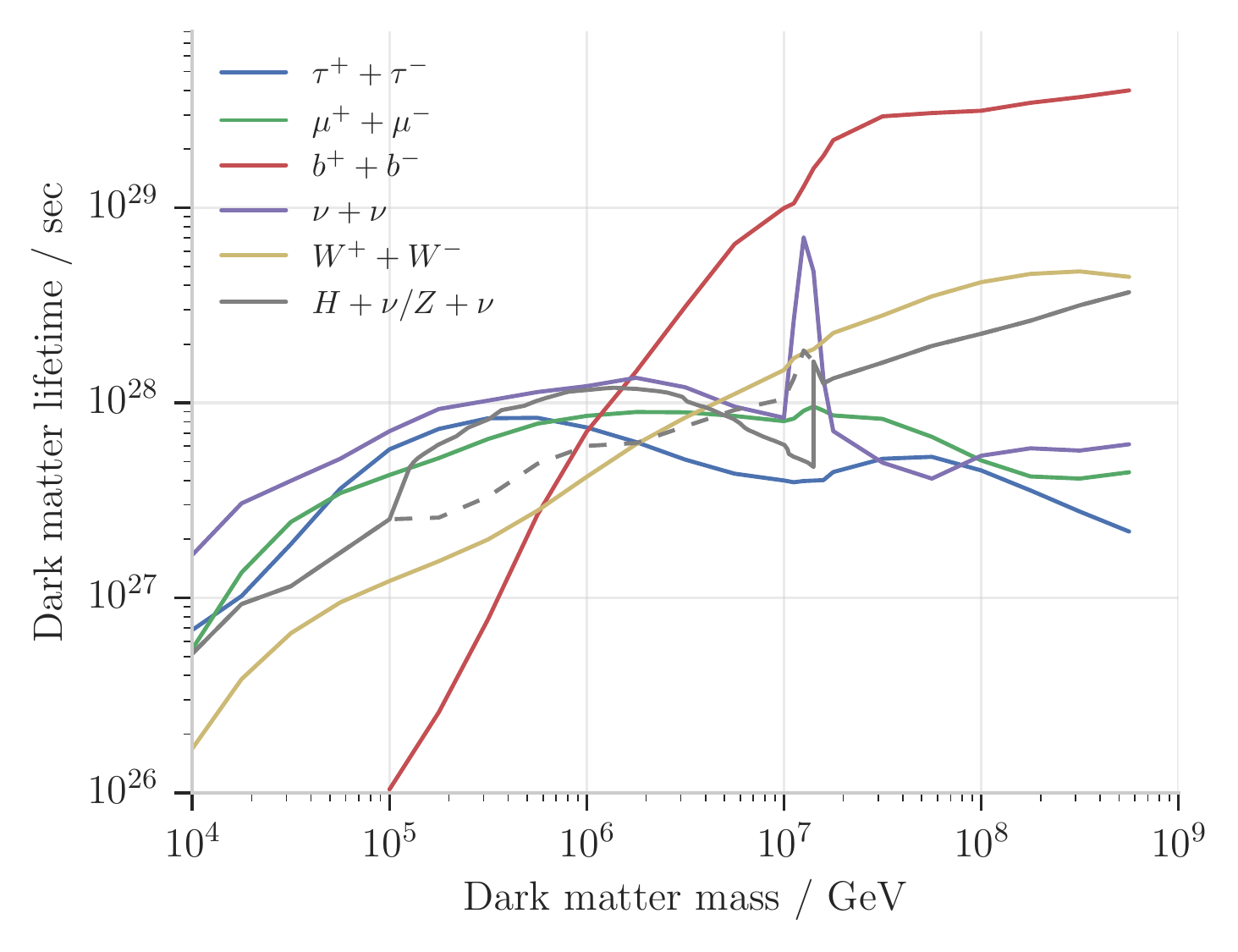} 
\end{center}
\caption{[Left] Upper limits on the DM annihilation cross section from the observation of neutrinos by ANTARES, IceCube, and Super-K (SK).
[Right] Lower limits on the DM lifetime from the observation of neutrinos by IceCube. 
Figures from Ref.~\cite{Super-Kamiokande:2020sgt} and Ref.~\cite{IceCube:2018tkk}, respectively.}
\label{fig:nu_spectra}
\end{figure*}

The particle nature of DM can be indirectly understood by observing its annihilation and/or decay products such as photons, $e^\pm$, $p^\pm$, and  $\nu$, coming from the regions where DM would be densely populated, e.g. the Galactic Center (GC) and the Sun.
Since a neutrino propagates along an almost straight line from where it is produced due to its weakly interacting nature, the observation of DM-induced neutrino signals allows to infer the source point. 
Detection of neutrino signals requires a large volume detector due to its weak interaction.
As mentioned in the introduction, various large-volume neutrino experiments or telescopes are already in operation or planned.
Examples include Antarctic Impulsive Transient Antenna (ANITA)~\cite{ANITA:2008mzi}, Astronomy with a Neutrino Telescope and Abyss environmental RESearch (ANTARES)~\cite{ANTARES:2011hfw}, Deep Underground Neutrino Experiment (DUNE)~\cite{DUNE:2018mlo, DUNE:2018hrq}, Giant Radio Array for Neutrino Detection (GRAND)~\cite{GRAND:2018iaj},  IceCube~\cite{IceCube:2016zyt, IceCube-Gen2:2020qha}, Jiangmen Underground Neutrino Observatory (JUNO)~\cite{JUNO:2021vlw}, KM3NeT~\cite{KM3Net:2016zxf}, Pierre Auger Observatory~\cite{PierreAuger:2015eyc}, Probe of Extreme Multi-Messenger Astrophysics observatory (POEMMA)~\cite{POEMMA:2020ykm}, Radio Neutrino Observatory in Greenland (RNO-G)~\cite{RNO-G:2020rmc}, Super-Kamiokande (Super-K)~\cite{Super-Kamiokande:2002weg}, and Hyper-Kamiokande (Hyper-K)~\cite{Hyper-Kamiokande:2016srs, Hyper-Kamiokande:2018ofw}.
Due to the high capability and large volume of their detectors, these experiments are expected to explore a wider range of parameter space in models of DM producing neutrinos via DM annihilation or decay. 
For a given DM model, one can easily calculate the expected neutrino flux from DM annihilation or decay.
Then, the search strategy is to find an excess of neutrinos from, e.g. the GC direction compared to the expected atmospheric neutrino background. 
No excess of neutrinos has been observed yet, and hence provides upper limits on DM annihilation cross sections~\cite{ANTARES:2020leh, Super-Kamiokande:2020sgt, Guepin:2021ljb} (see the left panel of \autoref{fig:nu_spectra}) and DM decay widths~\cite{IceCube:2018tkk, Guepin:2021ljb} (see the right panel of \autoref{fig:nu_spectra}) depending on annihilation or decay channels.

\begin{figure*}
\begin{center}
\includegraphics[width=0.75\textwidth]{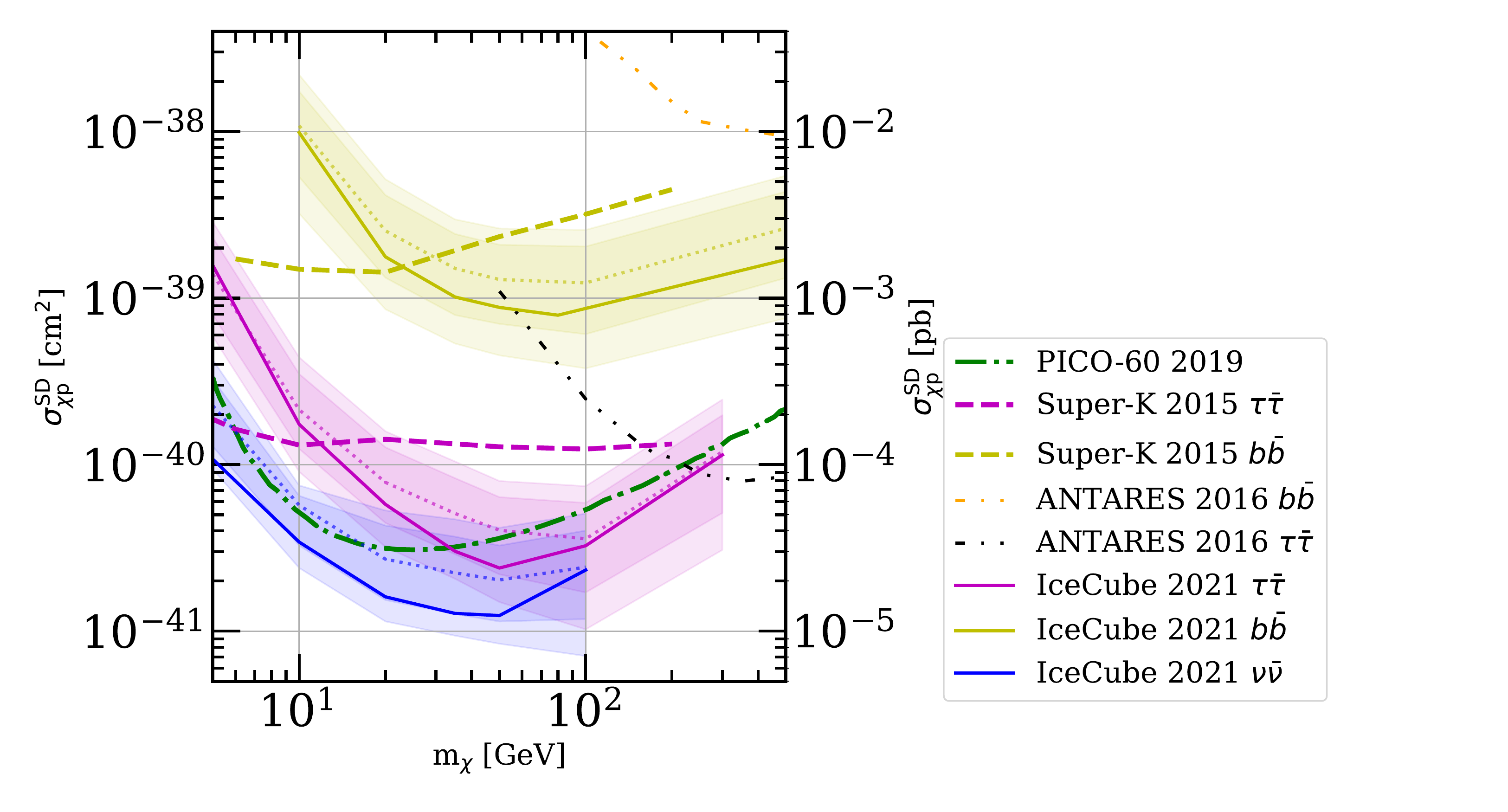} 
\end{center}
\caption{Upper limits on the spin-dependent DM-proton scattering cross section from neutrino signatures of DM annihilation in the Sun. 
Figure from Ref.~\cite{IceCube:2021xzo}.}
\label{fig:nu_spectra_sun}
\end{figure*}

DM can be accumulated inside astrophysical objects in the solar system such as the Earth~\cite{Krauss:1985aaa, Gould:1987ir, Gould:1987ww, Bruch:2009rp} and the Sun~\cite{Krauss:1985aaa, Gould:1987ir, Damour:1998vg, Bruch:2009rp, Zentner:2009is, Chen:2014oaa} through DM-nuclei and/or DM-dark-DM scatterings, and the captured DM may annihilate into SM particles in the Sun.
Thus, the Earth and the Sun can be good point-like sources of neutrino flux from DM annihilation with a relatively short distance from detectors compared to the GC.
Searches for an excess of neutrinos from the direction of the Sun or the center of the Earth over the known neutrino background also have been conducted. 
Upper limits on spin-independent and spin-dependent DM-nucleon scattering cross sections are now available~\cite{Super-Kamiokande:2015xms, ANTARES:2016xuh, ANTARES:2016bxz, IceCube:2021xzo} as shown in \autoref{fig:nu_spectra_sun}, since the DM capture processes rely on DM-nucleus scattering.
Interestingly, neutrino detectors provide more stringent limits on the spin-dependent DM-proton scattering cross section than those from DM direct-detection experiments depending on annihilation channels and DM masses.

\subsection{Boosted Dark Matter\label{sec:bdm}}

The novel phenomena of BDM generically arise in models beyond the minimal WIMP scenario, where a small relativistic component of DM~\cite{Huang:2013xfa, Agashe:2014yua, Berger:2014sqa, Kong:2014mia, Kim:2016zjx} can be produced and detected via its interactions with the SM particles.
The detection of BDM could be a smoking gun for DM discovery when it is challenging to detect the dominant component of cold DM.
Yet it requires new experimental strategies beyond conventional (cold relic) DM direct detection techniques which focus on low recoil energy (with the exception of very low-mass DM~\cite{Cherry:2015oca, Cui:2017ytb, Giudice:2017zke, McKeen:2018pbb, Kim:2020ipj}).
Due to the relatively small flux of BDM and the energetic states that are generally produced in the outcome of BDM interaction in detectors, large volume neutrino detectors are desirable and sensitive facilities for BDM searches.

In the earlier studies~\cite{Agashe:2014yua, Berger:2014sqa, Kong:2014mia}, BDM arises from models of two-component DM~\cite{Belanger:2011ww} which includes a cold component (say, $\psi$) as the dominant component of DM with very weak interaction with the SM, and a relativistic, lighter, secondary component (say, $\chi$) produced by the annihilation of $\psi$, which interacts with the SM states effectively. 
In the early universe, the $\psi\psi\rightarrow\chi\chi$ annihilation may dominate the thermal freeze-out of DM, thus determine DM relic abundance: the so-called {\it assisted freeze-out}~\cite{Belanger:2011ww}.
This serves as a new realization of WIMP miracle that is largely alleviated from strong experimental constraints on conventional WIMP DM~\cite{Belanger:2011ww, Agashe:2014yua, Berger:2014sqa, Kong:2014mia}.
In the present day, annihilation in DM-concentrated regions, such as the GC~\cite{Agashe:2014yua, Cherry:2015oca, Necib:2016aez, Alhazmi:2016qcs, Kim:2016zjx, Giudice:2017zke, Chatterjee:2018mej, Kim:2018veo, Aoki:2018gjf, McKeen:2018pbb, Kim:2019had, Kim:2020ipj, DeRoeck:2020ntj, Fornal:2020npv, Alhazmi:2020fju, Borah:2021jzu, Borah:2021yek}, the Sun~\cite{Berger:2014sqa, Kong:2014mia, Alhazmi:2016qcs, Berger:2019ttc} or dwarf galaxies~\cite{Necib:2016aez}, generates a BDM $\chi$ flux.
A BDM flux can also emerge from other scenarios of dark sectors, such as semi-annihilating DM~\cite{DEramo:2010keq, Berger:2014sqa, Chigusa:2020bgq, Toma:2021vlw}, self-annihilating DM~\cite{Carlson:1992fn, Hochberg:2014dra}, decaying DM~\cite{Bhattacharya:2014yha, Kopp:2015bfa, Bhattacharya:2016tma, Cui:2017ytb,  Heurtier:2019rkz}, DM-induced nucleon decay~\cite{Davoudiasl:2010am, Huang:2013xfa},  charged cosmic-ray acceleration~\cite{Bringmann:2018cvk, Ema:2018bih, Cappiello:2019qsw, Dent:2019krz, Wang:2019jtk, Ge:2020yuf, Cao:2020bwd, Jho:2020sku, Cho:2020mnc, Dent:2020syp, Bell:2021xff, Xia:2021vbz}, cosmic-ray neutrino acceleration~\cite{Jho:2021rmn, Das:2021lcr, Chao:2021orr}, astrophysical processes~\cite{Kouvaris:2015nsa, Hu:2016xas, An:2017ojc, Emken:2017hnp, Calabrese:2021src, Wang:2021jic} or inelastic collision of cosmic rays with the atmosphere~\cite{Alvey:2019zaa, Su:2020zny}.

As BDM arrives at a terrestrial neutrino detector, it can be detected via its interaction with electrons or hadrons, depending on the model specifics.
Extensive studies for signals from the BDM scattering off electrons have been conducted~\cite{Agashe:2014yua, Kong:2014mia, Necib:2016aez, Alhazmi:2016qcs, Kim:2016zjx, Super-Kamiokande:2017dch, Giudice:2017zke, Chatterjee:2018mej,  Kim:2018veo, COSINE-100:2018ged, Kim:2020ipj, DeRoeck:2020ntj, Cao:2020bwd, Alhazmi:2020fju, Chigusa:2020bgq, Chen:2021ifo}.
BDM scattering off hadrons, i.e.,~protons and neutrons, is more complex, and includes both (quasi-)elastic and (deep) inelastic modes.
The utilization of hadronic interactions could nevertheless be the discovery channel as shown in well-motivated models~\cite{Nelson:1989fx, He:1989mi, Carone:1994aa, Bailey:1994qv, FileviezPerez:2010gw, Graesser:2011vj, Batell:2014yra, Tulin:2014tya, Dobrescu:2014fca}.
Signatures of elastic BDM scattering off hardons have been investigated in e.g.~\cite{Berger:2014sqa, Cherry:2015oca, Kim:2016zjx, Berger:2019ttc, Kim:2020ipj, Wang:2021nbf, PandaX-II:2021kai, CDEX:2022fig}.
Signatures of inelastic BDM (iBDM) feature multiple visible particles in the final state~\cite{Kim:2016zjx, Giudice:2017zke, Bell:2021xff}, as the BDM upscatters to a heavier dark-sector state which decays back to the BDM state together with additional SM particles.
Therefore, typical iBDM signals suffer less from  potential backgrounds since additional signatures can be handles to distinguish them from backgrounds, allowing for competitive sensitivity reaches at high-capability neutrino~\cite{Kim:2016zjx, Chatterjee:2018mej, Chen:2020oft, Kim:2020ipj, DeRoeck:2020ntj} and DM detectors~\cite{Giudice:2017zke, COSINE-100:2018ged, Kim:2020ipj, Bell:2021xff}.  

In the following, we review a few representative examples of BDM scenarios, including the BDM flux sources, interaction patterns, and detection prospects.

%

\subsubsection{Multi-component boosted dark matter\label{sec:twocomponentbdm}}

\noindent {\bf Annihilation}: One of the most well-known BDM scenario is the annihilating two-component DM model as proposed in~\cite{Belanger:2011ww, Agashe:2014yua}.
The dark sector consists of two different DM species whose stability is protected by unbroken separate symmetries such as ${\rm U}(1)'\otimes{\rm U}(1)''$~\cite{Belanger:2011ww}.
The heavier DM particle, say $\chi_0$, does not directly interact with the SM particles, while the lighter one, say $\chi_1$, has a direct coupling to the SM particles.
On the other hand, the sizable direct interaction between $\chi_0$ and $\chi_1$ is allowed.

In this scenario, $\chi_1$ from the $\chi_0$ annihilation, $\chi_0\bar{\chi}_0 \rightarrow \chi_1\bar{\chi}_1$, obtains a sizable boost factor which is simply given by the mass ratio between $\chi_0$ and $\chi_1$, i.e., $\gamma_1=m_0/m_1$.
Assuming that $\chi_0$ follows the Navarro-Frenk-White (NFW) profile~\cite{Navarro:1995iw,Navarro:1996gj}, we estimate the expected flux of boosted $\chi_1$ near the Earth from all sky~\cite{Agashe:2014yua, Kim:2018veo}:
\begin{eqnarray}
\frac{d\Phi_1}{dE_1} &=& 
\frac{1}{4\pi}\int d\Omega \int_{\rm l.o.s.} ds \langle \sigma v \rangle _{\chi_0\bar{\chi}_0 \rightarrow \chi_1\bar{\chi}_1} \frac{dN_1}{dE_1} \left(\frac{\rho_0(r(s,\theta))}{m_0}\right)^2  \nonumber \\
&=& 3.2\times 10^{-4}~{\rm cm}^{-2}{\rm s}^{-1}\times \left(\frac{\langle \sigma v \rangle _{\chi_0\bar{\chi}_0 \rightarrow \chi_1\bar{\chi}_1}}{5\times 10^{-26}~{\rm cm}^3\,{\rm s}^{-1}} \right) \left( \frac{{\rm GeV}}{m_0}\right)^2 \frac{dN_1}{dE_1}\,, \label{eq:fluxAnnihilation}
\end{eqnarray}
where $\langle \sigma v \rangle _{\chi_0\bar{\chi}_0 \rightarrow \chi_1\bar{\chi}_1}$ is the velocity-averaged annihilation cross section in the universe today, $\rho_0$ is the $\chi_0$ density in our galaxy as a function of the distance $r$ to the GC, $s$ is the line-of-sight distance to the earth, $\theta$ is its angular direction relative to the earth-GC axis, and $\Omega$ is the solid angle.
The annihilation of $\chi_0$ yields a pair of mono-energetic $\chi_1$ particles whose differential energy spectrum is given by
\begin{equation}
\frac{dN_1}{dE_1}=2\delta(E_1 - m_0)\,.
\end{equation}
We assume here that $\chi_0$ is the dominant DM species, and $\chi_0$ and its antiparticle $\bar{\chi}_0$ are distinguishable and thus their fractions are same.
Thus, an additional pre-factor 1/2 is needed for the indistinguishable case. 
As variations of the annihilating galactic BDM, solar captured BDM scenarios~\cite{Berger:2014sqa, Kong:2014mia} have been explored where heavier (dominant) DM species can be efficiently captured in the Sun and annihilate into BDM.
For the solar captured BDM, self-interaction of the heavier DM greatly enhances the capture rate in the Sun and results in promising BDM fluxes at current and future experiments~\cite{Kong:2014mia}.

Interestingly, multi-component DM scenarios can show distinctive cosmological dynamics since the thermal evolution of the sub-component DM is significantly affected by the sizable self-scattering that is naturally realized for sub-GeV masses~\cite{Kamada:2021muh}.  
Consequently, warm DM constraints from the Lyman-$\alpha$ forest data and the number of satellite galaxies in the Milky Way can be significant.

\medskip

\noindent {\bf Decay}: 
Another BDM scenario with two-component DM is the decay of a heavy DM species to another lighter DM species~\cite{Bhattacharya:2014yha, Kopp:2015bfa, Heurtier:2019rkz}.
The dark sector have two DM particles: a heavy scalar DM $\phi$ with mass $m_\phi$ and a light fermionic DM $\chi$ with mass $m_\chi$.
The heavier component $\phi$ comprises a majority of the DM in the universe and decays into a pair of $\chi$'s with a lifetime $\tau_\phi$ greater than the age of the universe.
The lighter species $\chi$ directly interacts with SM particles and may deposit an observable amount of energy to electrons and/or nuclei in terrestrial detectors.

The $\chi$ pair from the decay of $\phi$ are mono-energetic and obtain a large boost factor $\gamma_\chi=E_\chi/m_\chi=m_\phi/2m_\chi$ due to the mass gap between the two DM species.
The flux of boosted $\chi$ is composed of a galactic component $\Phi_\chi^{\rm G}$ and an extragalactic component $\Phi_\chi^{\rm EG}$~\cite{Esmaili:2012us, Bai:2013nga, Bhattacharya:2014yha}:
\begin{equation}
\frac{d\Phi_\chi}{dE_\chi} = \frac{d\Phi_\chi^{\rm G}}{dE_\chi} + \frac{d\Phi_\chi^{\rm EG}}{dE_\chi}\,.   \label{eq:fluxDecay}
\end{equation}
The galactic component is 
\bea
\frac{d\Phi_\chi^{\rm G}}{dE_\chi} &=& \frac{1}{4\pi}\int d\Omega \int_{\rm l.o.s.} ds \frac{1}{\tau_\phi} \frac{dN_\chi}{dE_\chi} \left(\frac{\rho_\phi(r(s,\theta))}{m_\phi}\right)  \nonumber \\
&=& 2.1\times 10^{-4}~{\rm cm}^{-2}{\rm s}^{-1}\times \left(\frac{10^{26} {\rm sec}}{\tau_\phi} \right) \left( \frac{{\rm GeV}}{m_\phi}\right) \frac{dN_\chi}{dE_\chi}\,, \label{eq:fluxDecayG}
\eea
where $\rho_\phi$ is the $\phi$ density in the galaxy and the energy spectrum is given by
\begin{equation}
\frac{dN_\chi}{dE_\chi}=2\delta\left(E_\chi - \frac{m_\phi}{2}\right)\,.
\end{equation}
The extragalactic component is    
\bea
\frac{d\Phi_\chi^{\rm EG}}{dE_\chi} = \frac{\Omega_{\rm \phi}\rho_c}{m_\phi \tau_\phi} \int_0^\infty dz [(1+z)E_\chi] \frac{1}{H(z)} \frac{dN_\chi}{dE_\chi} \,, \label{eq:fluxDecayEG}
\eea
where $\Omega_{\rm \phi}$ is the $\phi$ relic density, $\rho_c$ is the critical density of the universe, and $H(z)\simeq H_0 \sqrt{\Omega_\Lambda + \Omega_{\rm m} (1+z)^3}$ is the Hubble expansion rate as a function of redshift $z$.

\subsubsection{Semi-annihilation}

BDM can also arise in models where the DM annihilation final state contain DM particle itself. This include, e.g., $3\rightarrow2$ self-annihilating DM \cite{Carlson:1992fn,Hochberg:2014dra}, and semi-annihilating DM \cite{DEramo:2010keq,Berger:2014sqa} where two DM particles $\chi$ annihilate into one DM particle along with a lighter particle $\psi$. Due to the kinematics in these scenarios, the final state DM becomes relativistic at production. The semi-annihilation case is most studied in the context of BDM, and will be our focus here. 

Semi-annihilation is generically motivated from models where DM particle $\chi$ is charged under a $Z_3$ symmetry, which protects $\chi$ from decay. The DM $\chi$ can either be a scalar or a fermion. Among the final states of the annihilation process $\chi\chi\rightarrow \chi \psi$, $\psi$ is not charged under $Z_3$, and thus without symmetry protection it is unstable and can decay to SM states. The decay products and the lifetime of $\psi$ are highly model dependent, but do not affect the essentials for BDM phenomena. 

Considering the initial states of DM particles being almost at rest (relevant to thermal freezeout and current-day detection), the Lorentz boost factor of DM particle in the final state is
\begin{equation}
  \label{eq:BoostFactor}
  \gamma_{\chi}=\frac{(5m_{\chi}^2-m_{\psi}^2)}{4m_\chi^2},
\end{equation}
 where $\gamma_\chi$ ranges from
1 to 1.25, depending on $m_\psi$. In the general case of $m_{\psi}\ll m_\chi$, the maximal boost, 1.25, is reached. As shown in \cite{Berger:2014sqa},  a boost factor around 1.25 is in the
\textit{sweet-spot} for detecting BDM at Cherenkov neutrino detectors via hadronic channels, as it allows the outgoing nucleon to pass Cherenkov limit, without entering the messy deep inelastic regime.

BDM from semi-annihilation can be sourced in various astrophysical objects including the GC. A particularly interesting case is BDM from the Sun through solar capture of the annihilating DM, which was studied in \cite{Berger:2014sqa}. In this scenario, the scattering cross-section between DM particle and nucleon determines both DM solar capture rate and the
interaction event rate as a BDM particle passes through a terrestrial detector, which reduces the number of free parameters in the model. In addition, the BDM flux is expected to be larger due to the closer distance between the Sun and the Earth, AU.
The flux in this solar capture case can be written as
\begin{equation}
  \label{eq:flux}
  \Phi = \frac{A N^2}{4 \pi {\rm AU}^2},
\end{equation}
where $N$ is the DM number in the Sun, $A N^2$ relates to the annihilation rate of
DM captured in the Sun, $\Gamma_A$ is given by $\Gamma_A=\frac{1}{2}A N^2$. The annihilation rate $A$, in turn, is effectively given by the capture rate $C=AN^2$ in the parameter space of most interest, where the equilibrium between annihilation and capture is reached in the Sun. Other effects such as evaporation and rescattering are also taken into account in \cite{Berger:2014sqa}. Various possibilities of BDM species (fermion or scalar) and interaction operators (spin and velocity dependence) relevant for detection were considered as well.


\subsubsection{Charged-cosmic-ray-induced boosted dark matter}
Another well-studied BDM scenario is DM boosted by energetic charged cosmic-ray particles such as electron, proton, and helium as long as interactions between DM and cosmic-ray particles are allowed~\cite{Bringmann:2018cvk, Ema:2018bih, Cappiello:2019qsw, Dent:2019krz, Jho:2020sku, Cho:2020mnc}. 
Energetic charged-cosmic-rays scattering off non-relativistic relic DM particles in the galaxy can induce a subdominant (semi-)relativistic DM component.

For a charged-cosmic ray particle $i$, the flux of boosted $\chi$ can be obtained by convolving a DM-$i$ differential scattering cross section with the cosmic ray $i$ flux~\cite{Bringmann:2018cvk}:
\bea
\frac{d\Phi_\chi}{dK_\chi} = \frac{1}{4\pi} \int d\Omega \int_{\rm l.o.s.} ds \left(\frac{\rho_\chi(r(s,\theta))}{m_\chi}\right) \int_{K_i^{\rm min}}^\infty dK_i \frac{d\sigma_{\chi i}(K_i)}{dK_\chi} \frac{d\Phi_i^{\rm LIS}}{dK_i}\,, \label{eq:fluxChargedCR}
\eea
where $\rho_\chi$ is the relic density of $\chi$ in the galaxy, $d\Phi_i^{\rm LIS}/dK_i$ is the local interstellar  differential flux of the charged-cosmic ray $i$~\cite{Bisschoff:2019lne}, and $K_i^{\rm min}$ is the minimum kinetic energy of the cosmic ray particle $i$ to produce DM with a kinetic energy $K_\chi$ after the collision, which is given by
\bea
K_i^{\rm min} = \left( \frac{K_\chi}{2} - m_i \right) \left( 1 \pm \sqrt{1 + \frac{2K_\chi}{m_\chi} \frac{(m_i+m_\chi)^2}{(2m_i-K_\chi)^2}} \right)\,,
\eea
where the plus and minus signs correspond to $K_\chi > 2m_i$ and $K_\chi < 2m_i$, respectively.
This mechanism yields a continuous energy spectrum of BDM.

\subsubsection{Cosmic-ray neutrino-induced boosted dark matter}

\noindent {\bf Stellar neutrino}: 
Another class of DM boosting mechanism, cosmic-ray neutrino-BDM ($\nu$BDM), has been suggested in Ref.~\cite{Jho:2021rmn}.
A huge number of cosmic-ray neutrinos from stars in the galaxy can scatter off galactic DM particles and induce energetic DM components.
Neutrinos from stars can very efficiently boost light DM ($\lesssim 10$ MeV), and $\nu$BDM can dominate the whole cosmic-ray-induced BDM when DM-neutrino interaction is as strong as DM-electron/nucleus interaction.

The boosted $\chi$ flux by neutrinos from a single Sun-like star is given by~\cite{Jho:2021rmn}
\begin{eqnarray}
\frac{d\Phi^{(1)}_\chi(\overrightarrow{y})}{dK_\chi}
\simeq \frac{1}{8\pi^2} \left( \tilde{f}_1 \frac{d\dot{N}^{\rm Sun}_\nu}{dK_\nu} \right)
&&\int d^3 \overrightarrow{z} \frac{ \rho_\chi(|\overrightarrow{z}|)}{m_\chi} \frac{1}{|\overrightarrow{x}-\overrightarrow{z}|^2} \nonumber \\
&&\times
\frac{1}{\sin\bar{\theta}_0}\frac{1}{|\overrightarrow{z}-\overrightarrow{y}|^2}
\times \exp {\left(-\frac{|\overrightarrow{z}-\overrightarrow{y}|}{d_\nu} \right)} \nonumber \\
&& \times \left( \left.
\frac{dK_\nu}{d\bar{\theta}}\right\vert_{\bar{\theta}=\bar{\theta}_0}
 \right)
 \left(
 \left. \frac{d\sigma_{\chi\nu}}{dK_\chi}\right\vert_{\bar{\theta}=\bar{\theta}_0}
 \right)  \,,
\label{eq:fluxNuOneStar}
\end{eqnarray}
where $\overrightarrow{x},\overrightarrow{y}$, and $\overrightarrow{z}$ respectively represent the positions of the Earth, a star, and a halo DM centered on the GC, $d\dot{N}^{\rm Sun}_\nu/dK_\nu$ is the neutrino emission rate for a Sun-like star~\cite{Bahcall:2004pz, Billard:2013qya, Vitagliano:2017odj}, and $\tilde{f}_1$ reflects the variances of stellar properties from the Sun.
The attenuation of the neutrino flux due to propagation is determined by the exponential function in the second line where $d_\nu$ is the mean free path.
The last line takes into account the fraction that the scattering angle of $\chi$, $\bar{\theta}$, coincides with the direction to the Earth $\bar{\theta}_0$, which is determined by $K_\nu$ and $K_\chi$ via kinematic relations where $\sigma_{\chi\nu}$ is the DM-neutrino scattering cross-section.
By convolving Eq.~(\ref{eq:fluxNuOneStar}) with entire stellar distribution $n_{\rm star}(\overrightarrow{y})$ in the galaxy~\cite{deJong:2009iq}, one can obtain the total $\nu$BDM flux~\cite{Jho:2021rmn}:
\begin{eqnarray} 
\frac{d\Phi_\chi}{dK_\chi} = \int d^3 \overrightarrow{y} n_{\rm star}(\overrightarrow{y}) \frac{d\Phi^{(1)}_\chi(\overrightarrow{y})}{dK_\chi}\,.
\label{eq:fluxNuTotal}
\end{eqnarray}
The $\nu$BDM mechanism yields a broad energy spectrum of BDM.
The galactic DM particles can also get boosted and have semi-relativistic velocities by scattering with the diffuse supernova neutrino background (DSNB)~\cite{Das:2021lcr} and the neutrinos from evaporation of primordial black holes~\cite{Chao:2021orr}.

\noindent {\bf Supernova neutrino}: 
Allowing for neutrino-DM interactions also opens up the possibility of DM being boosted by the relic neutrinos present from all massive stars going core-collapse supernova (SN) throughout the history of star-formation.
This DSNB is ubiquitous, and consists of a swathe of MeV neutrinos (and anti-neutrinos) of all flavors arriving at the Earth isotropically.
As a result, a fraction of the DM in the Milky Way halo can undergo scattering with the DSNB, and get accelerated towards the Earth.
This was studied in Ref.~\cite{Das:2021lcr}.

The flux of such supernova BDM can be computed as follows.
The DSNB flux at the Earth requires knowledge on the rate of core-collapse SN happening in the universe $(R_{\rm CCSN})$, as well as the underlying cosmological model, encapsulated through the Hubble paramter $H(z)$, and the neutrino spectra from a SN, $F_\nu(E)$, usually treated as a pinched Fermi-Dirac spectra.
With this, the DSNB can be estimated as\,\cite{Beacom:2010kk},
\begin{equation}\label{eq:DSNB}
\Phi_{\nu}(E_\nu)=\int_0^{z_{\rm max}}\frac{dz}{H(z)}R_{\rm CCSN} F_\nu[E_\nu\,(1+z)]\,,
\end{equation}
where $z_{\rm max}\sim 5$ denotes the maximum redshift of star-formation.
Assuming an energy transfer of $T_\chi$ to the DM, the 
boosted DM flux is,
\begin{equation}\label{eq:bdm_flux2}
	\frac{d\phi_\chi}{dT_\chi} = \int dE_\nu \frac{d\phi_\chi}{dE_\nu} \frac{1}{T_\chi^\mathrm{max}(E_\nu)} \Theta\left[T_\chi^\mathrm{max}(E_\nu)-T_\chi\right]\,
\end{equation}
where $T_\chi^\mathrm{max}(E_\nu)$ is the maximum energy transferred to the DM by a neutrino of energy $E_\nu$, $\Theta[\ldots]$ is the Heaviside theta function, and the upscattered DM flux as a function of neutrino energy is given by
\begin{equation}\label{eq:bdm_flux1}
	\frac{d\phi_\chi}{dE_\nu} =  D_\mathrm{halo} \Phi_\nu(E_\nu) \frac{\sigma_{\chi\nu}}{m_\chi}\,.
\end{equation}
Here $D_\mathrm{halo}$ denotes the line-of-sight and angular integrals over the DM density in the halo (which is taken to follow the NFW profile), while $\sigma_{\chi\nu}$ denotes the DM-neutrino interaction cross-section.

\subsubsection{Searches for boosted dark matter\label{sec:BDMsearches}}

\begin{figure}
\centering
\includegraphics[width=0.48\linewidth]{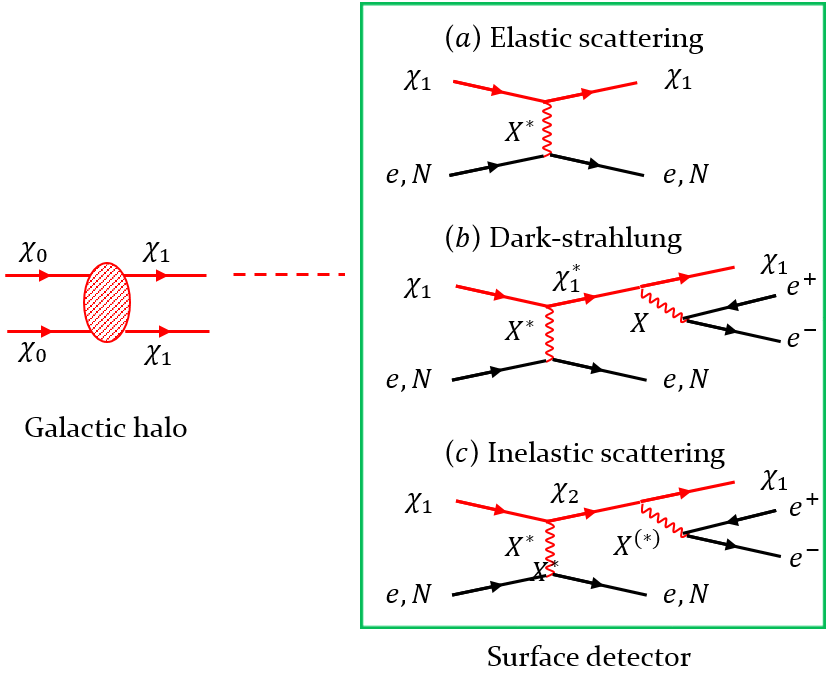} 
\caption{\label{fig:ES} 
Example BDM signal processes. 
}
\end{figure}

Large-volume neutrino experiments installed deep underground such as DUNE and Super-K/Hyper-K have been considered in order to probe BDM signals from dark-sector annihilation processes~\cite{Super-Kamiokande:2017dch, DUNE:2020ypp, DUNE:2020fgq}.
Since the cosmic-ray fluxes are quite small in deep underground, the main background is the neutral current scattering of atmospheric neutrinos, which can be reduced by using the {\it directional information} of BDM coming from specific areas such as the GC, the Sun, and dwarf spheroidal galaxies~\cite{Agashe:2014yua, Berger:2014sqa, Kong:2014mia, Necib:2016aez, Alhazmi:2016qcs, Super-Kamiokande:2017dch, Berger:2019ttc} or (almost completely) rejected in scenarios allowing for signals with {\it secondary signatures} in addition to the primary target recoil, e.g., iBDM~\cite{Kim:2016zjx, Giudice:2017zke, COSINE-100:2018ged, Kim:2020ipj, DeRoeck:2020ntj} and dark-strahlung (dark gauge boson bremsstrahlung from BDM)~\cite{Kim:2019had} (see \autoref{fig:ES}).
Depending on the model details, conventional DM direct detection experiments (preferably ton-scale and above) can be sensitive to the BDM signals.
Example studies include the BDM searches in the nuclear recoil channel~\cite{Cherry:2015oca} and in the electron recoil channel~\cite{Giudice:2017zke}. 
The COSINE-100 collaboration performed an iBDM search in the electron recoil channel~\cite{COSINE-100:2018ged}, and studies~\cite{Fornal:2020npv, Jho:2020sku, Alhazmi:2020fju} showed that BDM can explain the electron-recoil excess reported by the XENON1T collaboration~\cite{XENON:2020rca}.
The excessive events at TeV--PeV energies observed by the IceCube collaboration and at EeV energies by the ANITA collaboration can be explained by the scattering of highly boosted DM particles from the decay of a superheavy DM candidate~\cite{Bhattacharya:2014yha, Kopp:2015bfa, Heurtier:2019rkz}. 

By contrast, surface-based neutrino experiments such as the Short-Baseline Neutrino program (SBN) and the NuMI Off-axis $\nu_e$ Appearance (NO$\nu$A) face an enormous amount of cosmic-ray-induced background.
It is also possible to utilize features resulting from {\it unique event topologies} of the BDM signal to reject the cosmic-ray-induced background, for example, in the cases of iBDM~\cite{Kim:2016zjx} and dark-strahlung~\cite{Kim:2019had}.
The number of cosmic-ray-induced events being inseparable from the iBDM signals at ProtoDUNE are conservatively estimated to be smaller than $100~{\rm yr}^{-1}$~\cite{Chatterjee:2018mej}.
On the other hand, it is non-trivial to search for elastically scattering BDM signals that are featureless hence easily mimicked by typical backgrounds mentioned above.
In order to resolve this problem, the idea of {\it Earth Shielding}~\cite{Kim:2018veo} has been proposed.
The key observation is that the cosmic rays cannot penetrate the Earth unlike the BDM, and hence we can take only {\it upward-going} signals when a detector becomes located in the opposite side of the BDM source.
In the case that BDM comes from the GC, a simple geometric consideration suggests that the SBN detectors at Fermilab and the ProtoDUNE detectors at CERN can use $\sim 65$\% and $\sim70$\% of the signal flux incident for a year, respectively, without suffering from the cosmic-ray-induced backgrounds.

\begin{figure}[t]
    \centering
    \includegraphics[width=0.47\linewidth]{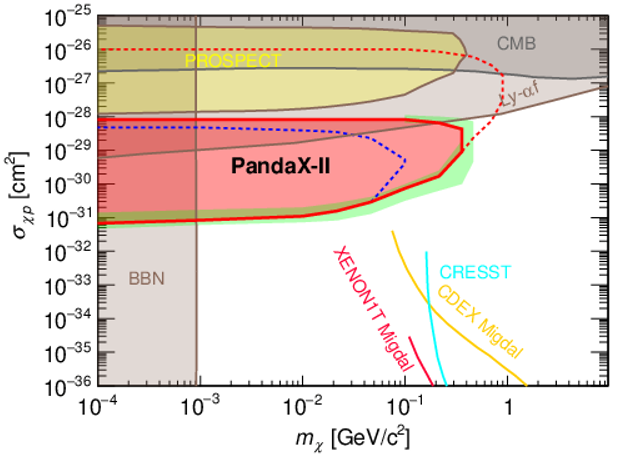}
    \includegraphics[width=0.49\linewidth]{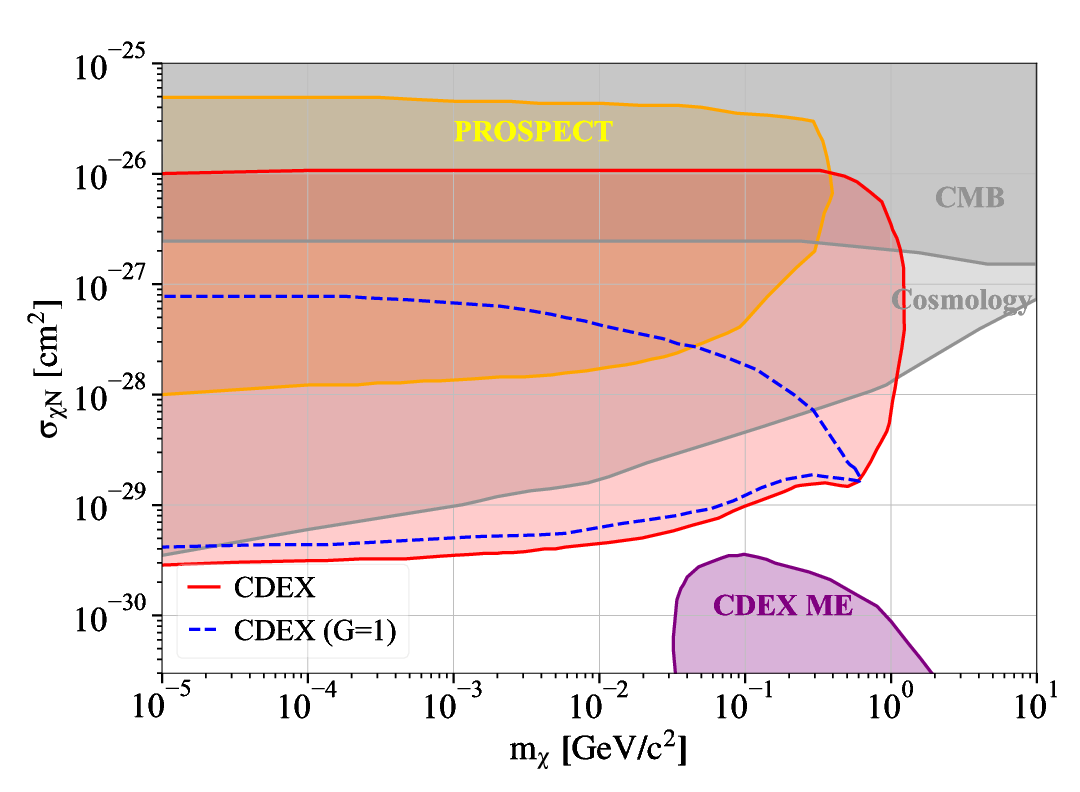}
    \caption{The limits of light DM boosted by cosmic protons/helium that are reported by the PandaX-II (left) and CDEX (right) Collaborations. Figures from~\cite{PandaX-II:2021kai} and~\cite{CDEX:2022fig}, respectively. }
    \label{fig:crbdmlimits}
\end{figure}

DM particles can be boosted and acquire relativistic or semi-relativistic speeds via scattering processes with cosmic rays such as protons, helium, electrons, and neutrinos, which induce detectable recoil signals over the thresholds of large-volume terrestrial detectors, efficiently for sub-GeV dark matter.
Cosmic-ray-induced BDM provides substantially-extended exclusion regions of parameter space based-on the existing data from experiments such as Borexino, CDEX, Daya Bay, KamLAND, MiniBooNE, Super-K, PandaX, and XENON1T for DM-nucleon/electron/neutrino elastic scattering~\cite{Bringmann:2018cvk, Ema:2018bih, Cappiello:2019qsw, Dent:2019krz, Cao:2020bwd, Jho:2020sku, Cho:2020mnc, Dent:2020syp, Jho:2021rmn, Das:2021lcr, Chao:2021orr};
moreover, projected sensitivity regions for future experiments such as DUNE, Hyper-K, JUNO, and XENONnT~\cite{Ema:2018bih, Cappiello:2019qsw, Dent:2020syp, Jho:2021rmn}.
Interestingly, cosmic-ray-induced BDM can account for the excessive low-energy electron-recoil events observed by the XENON1T experiment if it interacts with electrons (and neutrinos)~\cite{Jho:2020sku, Jho:2021rmn, Das:2021lcr, Chao:2021orr}.
Recently, new constraints on light DM boosted by cosmic protons/helium have been reported from the PandaX-II Collaboration~\cite{PandaX-II:2021kai} and the CDEX Collaboration~\cite{CDEX:2022fig}, as displayed in~\autoref{fig:crbdmlimits}.

\subsubsection{Boosted axion-like particles\label{sec:BDMaxion}}
Axion-like particles (ALPs) are compelling dark matter candidates that attracted rising interest in recent years, as an alternative to WIMP miracle. 
Most searches for cosmic ALPs so far focused on their cold relics as (a component of) dark matter. 
However, there are many well-motivated sources producing a cosmic background of energetic axions that remain relativistic today. With effective interaction with the SM particles, such ``boosted” axions can be detected by terrestrial experiments. Such a scenario resonates with the original idea of boosted DM (BDM) which has been well explored and reviewed in the earlier sections in this white paper. Nevertheless, the phenomenology and search strategies for the boosted ALPs and original BDM are distinct from each other. BDM is typically thermal WIMP-like particles generated from processes such as annihilation or decay of DM, and its detection is based on the scattering process. In contrast, boosted ALPs are very light and thus dark-radiation-like, with very weakly interactions, and thus would be detected by absorption processes such as via inverse Primakoff effect in the presence of axion-photon coupling. Large volume neutrino detectors are well-suited for ALPs of O(MeV) or higher energy (see \cite{Cui:2017ytb,Dror:2021nyr} for the possibility of detecting lower energy boosted ALPs with other types of experiments). 
 The recent work \cite{Cui:2022owf} takes ALPs from DM decay as an example, and demonstrates that current/upcoming neutrino experiments such as Super- Kamiokande (SK), Hyper-Kamiokande (HK), DUNE, JUNO, and IceCube can be sensitive to such energetic axion relics. The characteristic signature is the mono-energetic single photon signal from axion absorption induced by the axion-photon coupling. 
Among these experiments, both DUNE and JUNO are capable of probing $g_{a\gamma\gamma}$ values below existing astrophysical limits. In comparison, for the Cherenkov detectors SK and HK, the mono-$\gamma$ signal is subject to background, which yields a less competitive sensitivity relative to DUNE and JUNO. This new avenue potentially cover parameter ranges complementary to existing axion searches and provides new opportunities for dark matter related BSM discovery with neutrino facilities.

\subsection{Explosive Slow-Moving Dark Matter}
\label{sec:explosive}
In neutrino experiments, slow-moving DM scattering can lead to a striking signal if there is a mass difference between the initial and final state particles, causing an {\em explosion} in the detector. Classes of DM processes of this type are nuclear destruction~\cite{Davoudiasl:2010am,Davoudiasl:2011fj,Huang:2013xfa}, self destruction~\cite{Grossman:2017qzw,Geller:2020zhq}, and fermionic absorption~\cite{Kile:2009nn,Dror:2019onn,Dror:2019dib,Dror:2020czw,Ge:2022ius}.\footnote{Closely related examples include bosonic absorption~\cite{An:2013yua,An:2014twa,Bloch:2016sjj}, where the medium absorbs a boson, and exothermic processes~\cite{Graham:2010ca}, where there is a mass difference between the initial or final dark sector particle. However, in both these cases, the kinematics are such that a negligible signal will be detected in neutrino experiments.} These cases are depicted in \autoref{fig:processes} where visible-sector particles are illustrated in blue, dark-sector particles are labeled by $ \chi $ and shown in black, while the underlying interaction is depicted as a gray circle. Published work has focused on $ 2 \rightarrow 2 $ scattering processes, and we continue to do so in this white paper. If one allows for additional final state particles, the options are wider, but the phenomenology will be similar.

DM capable of inducing explosive interactions has a few unifying features:
\begin{enumerate} 
\item {\bf Potential Instability:} For explosive processes, the incoming DM species is not one of the outgoing states. This leads to the generic expectation that DM is not absolutely stable, a characteristic strongly constrained by DM indirect detection searches. This bound suggests specific types of DM interaction and masses. 
\item {\bf Mass Energy Deposition:} The kinematics of explosive scatterings are such that the energy deposited in the detector is set by mass energy, not kinetic energy. The corresponding high-energy events are the key feature that enables detecting such DM in higher threshold detectors within neutrino experiments. 
\item {\bf Baryon or Lepton Number:} Explosive DM often carries baryon or lepton number (or even both). This suggests a possible connection to the observed matter-antimatter asymmetry of the universe.
\end{enumerate}
We will see these features making a recurrent appearance in each of the cases, which we now describe in detail.  

\definecolor{c1}{HTML}{003262} 
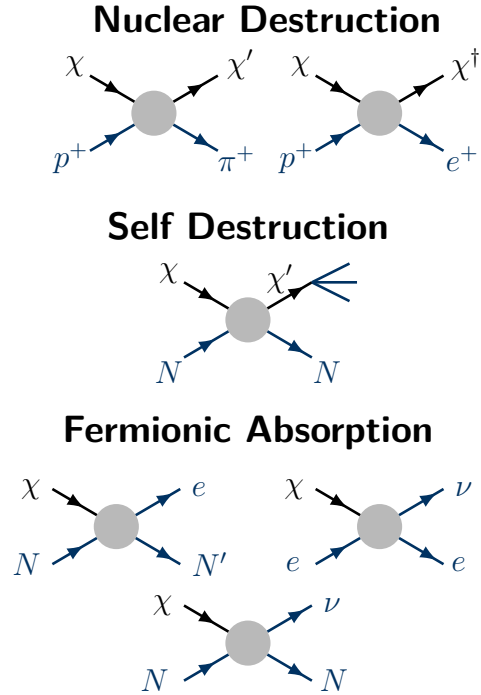
\begin{figure}
\begin{center} \begin{tikzpicture} [f/.style={draw,decoration={markings,mark=at position #1 with {\arrow[very thick]{latex}}},postaction={decorate},node contents=#1},
    f/.default=.6,
    fb/.style={draw,decoration={markings,mark=at position #1 with {\arrowreversed[very thick]{latex}}},postaction={decorate},node contents=#1},
    fb/.default=.4,line width=1
]

\coordinate[] (C1A) at (0,0);
\node at ($ (C1A)+(1.5,1.25) $) {\large \bf Nuclear Destruction};
\draw[f =0.5,c1] ($ (C1A)+(-0.85,-0.5) $) node[xshift=-0.25cm,yshift=-0.1cm] {$ p ^+   $} -- (C1A);
\draw[f =0.5] ($ (C1A)+(-0.85,0.5) $) node[xshift=-0.2cm,yshift=0.125cm] {$ \chi   $} -- (C1A)   ;
\draw[f =.8] (C1A) --++ (.85,.5) node[xshift=0.3cm,yshift=0.15cm] {$ \chi '   $} coordinate(A);
\draw[f =0.9,c1] (C1A) --++  (.85,-.5) node[xshift=0.25cm,yshift=-0.1cm]{$ \pi ^+   $};
\draw[preaction={fill=white}, fill=gray,fill opacity=0.55,draw=none] (C1A) circle (0.3) ;

\coordinate[] (C1B) at ($ (C1A) +(3.,0) $);
\draw[f =0.5,c1] ($ (C1B)+(-0.85,-0.5) $) node[xshift=-0.25cm,yshift=-0.1cm] {$ p ^+   $} -- (C1B);
\draw[f =.5] ($ (C1B)+(-0.85,0.5) $) node[xshift=-0.2cm,yshift=0.125cm] {$ \chi   $} -- (C1B)   ;
\draw[f =.8] (C1B) --++ (.85,.5) node[xshift=0.3cm,yshift=0.15cm] {$ \chi  ^\dagger   $} coordinate(A);
\draw[f =0.9,c1] (C1B) --++  (.85,-.5) node[xshift=0.25cm,yshift=-0.1cm]{$ e ^+   $};
\draw[preaction={fill=white}, fill=gray,fill opacity=0.55,draw=none] (C1B) circle (0.3) ;

\coordinate[] (C2) at ($(C1B)+ (-1.75,-2.75) $);
\node at ($ (C2)+(0,1.25) $) {\large \bf Self Destruction};
\draw[f =0.5,c1] ($ (C2)+(-0.85,-0.5) $) node[xshift=-0.2cm,yshift=-0.2cm] {$  N $} -- (C2);
\draw[f =0.5] ($ (C2)+(-0.85,0.5) $) node[xshift=-0.2cm,yshift=0.125cm] {$ \chi $} -- (C2)   ;
\draw[f =0.9] (C2) --++ (.85,.5) node[midway,xshift=-0.cm,yshift=0.3cm] {$ \chi ' $} coordinate(A);
\draw[f =0.9,c1] (C2) --++  (.85,-.5) node[xshift=0.2cm,yshift=-0.2cm]{$ N  $};
\draw[preaction={fill=white}, fill=gray,fill opacity=0.55,draw=none] (C2) circle (0.3) ; 
\draw[c1] (A) --+ (0.5,0.25);
\draw[c1] (A) --+ (0.6,0);
\draw[c1] (A) --+ (0.5,-0.25);

\coordinate[] (C3A) at ($ (C2)+(-1.75,-2.75) $);
\node at ($ (C3A)+(1.75,1.25) $) {\large \bf Fermionic Absorption};
\draw[f =0.5,c1] ($ (C3A)+(-0.85,-0.5) $) node[left] {$ N$} -- (C3A);
\draw[f =0.5] ($ (C3A)+(-0.85,0.5) $) node[left] {$ \chi $} -- (C3A)   ;
\draw[f =0.9,c1] (C3A) --++ (.85,.5) node[right] {$ e   $};
\draw[f =0.9,c1] (C3A) --++  (.85,-.5) node[right]{$ N ' $};
\draw[preaction={fill=white}, fill=gray,fill opacity=0.55,draw=none] (C3A) circle (0.3) ;

\coordinate[] (C3B) at ($ (C3A)+(3.5,0) $);
\draw[f =0.5,c1] ($ (C3B)+(-0.85,-0.5) $) node[left,xshift=-.05cm]{$ e $} -- (C3B);
\draw[f =0.5] ($ (C3B)+(-0.85,0.5) $) node[left] {$ \chi $} -- (C3B)   ;
\draw[f =0.9,c1] (C3B) --++ (.85,.5) node[right] {$ \nu   $};
\draw[f =0.9,c1] (C3B) --++  (.85,-.5) node[right,xshift=-.05cm]{$ e $};
\draw[preaction={fill=white}, fill=gray,fill opacity=0.55,draw=none] (C3B) circle (0.3) ; 

\coordinate[] (C3C) at ($ (C3A)+(1.75,-1.55) $);
\draw[f =0.5,c1] ($ (C3C)+(-0.85,-0.5) $) node[left,xshift=-.05cm]{$ N $} -- (C3C);
\draw[f =0.5] ($ (C3C)+(-0.85,0.5) $) node[left] {$ \chi $} -- (C3C)   ;
\draw[f =0.9,c1] (C3C) --++ (.85,.5) node[right] {$ \nu   $};
\draw[f =0.9,c1] (C3C) --++  (.85,-.5) node[right,xshift=-.05cm]{$ N $};
\draw[preaction={fill=white}, fill=gray,fill opacity=0.55,draw=none] (C3C) circle (0.3) ; 
  \end{tikzpicture}
\end{center}
\caption{Different processes in which slow-moving DM can induce an explosive signal in neutrino experiments through nuclear destruction ({\bf Top}), self-destruction ({\bf Middle}), and fermionic absorption ({\bf Bottom}). Visible sector particles are depicted in blue, while dark sector particles are depicted in black. The process is mediated by an interaction depicted as a gray circle.}
\label{fig:processes}
\end{figure}

\subsubsection{Nuclear destruction}
Nuclear destruction can take place when there are multiple states in the dark sector. One possible example is a Dirac fermion, $ ( \chi , \bar{\chi} ) $, and complex scalar, $ \Phi $, with an interaction with SM quarks. The operator and a potential corresponding experimental process is \cite{Davoudiasl:2010am,Davoudiasl:2011fj},
\begin{equation} 
{\cal L} \supset \frac{1}{ \Lambda ^3 } u ^c  d ^c d ^c  \chi ^c   \Phi +{\rm h.c.} ~\Rightarrow~ p ^+  + \chi \rightarrow  \pi  ^+  + \Phi \label{eq:LL}\,,
\end{equation} 
where we follow two-component notation denoting right-handed fields with a charge conjugation symbol ${} ^c $. 
We can see the three features mentioned above. Firstly, to conserve baryon number ($ B $) requires a non-vanishing baryon number in the dark sector, $ B _\chi +  B _{ \Phi } = - 1 $. Additionally, the DM is generically unstable unless $ m _\chi - m _\Phi \leq m _{ p} -  m _{ \pi }$. Lastly, the energy deposited in a Hydrogen-rich experiment with approximately-free movinag protons (e.g., Super-K or Borexino) is, $ \simeq  m _\chi + m _p - m _{ \Phi } - m _{ \pi }  $. This can be much larger than a {\rm MeV} and well above most neutrino experiment detector thresholds. Nuclear destruction of DM has a strong connection to baryogenesis; the operator in \autoref{eq:LL} can be a consequence of integrating out a heavy state which decays to both the visible and dark sector. If the interactions violate charge and parity, the decays result in an asymmetry in the visible-sector and dark-sector baryons simultaneously.

Nuclear destruction can also convert a baryon into a lepton through an interaction~\cite{Huang:2013xfa},
\begin{equation} 
{\cal L} \supset = \frac{1}{ \Lambda ^5 }  \chi     ^2 ( e ^c  u ^c  ) ( d  ^c  u ^c  ) +{\rm h.c.} ~ \Rightarrow ~ \chi + p ^+ \rightarrow  e ^+  +  \bar{\chi}  \,,
\end{equation} 
where $ \bar{ \chi} $ denotes the $ \chi $ antiparticle. Such a process has similar features to those described above.

\subsubsection{Self destruction}
If DM is part of a bound state (mediated by some additional long-range force), it may be excited to higher energy state through a collision with a target medium. If the higher energy state is unstable, it can rapidly decay depositing significant energy within the detector volume. One possible processes involves a Dirac fermion, $ \chi $, in a bound state, $ ( \chi \chi ) $, that interacts with a neutron, $ n $, and a new fermion, $ \psi $. The corresponding induced process is~\cite{Grossman:2017qzw}
\begin{equation} 
{\cal L} \supset \frac{ \chi ^2 ( \psi ^c n ) }{ \Lambda ^2 } +{\rm h.c.} ~ \Rightarrow ~ ( \chi \chi ) + n \rightarrow ( \chi \bar{\chi}   )  + \psi  \,.
\end{equation} 
In the reaction, the particle-particle bound state converts into a particle-antiparticle bound state which rapidly decays. We can again see the unifying features of explosive processes in self-destructing DM. Firstly, baryon number conservation implies $ 2 B _{ \chi } + B _{\psi ^c  } = - 1 $. Secondly, the energy deposited in the target medium is the mass energy, not the kinetic energy. Unlike the example above, however, the outgoing particle was chosen to be the antiparticle of $ \chi $ which effectively prevents the decay. 

Alternative DM models which exhibit self-destruction rely on $ \chi $ being trapped in high-angular momentum, $ ( \chi \bar{\chi} )  $, states~\cite{Grossman:2017qzw,Geller:2020zhq}. The sizable angular momentum can prevent the bound states from decaying on a cosmological timescale. However, these can be converted to a lower angular-momentum state through scattering and will subsequently decay into the ground state. 

\subsubsection{Fermionic absorption}
A final class of explosive DM scatterings is from the absorption of fermionic DM. In these scattering events, no dark particle is present in the final state; instead, a visible particle is ejected. Fermionic absorption operators are expected for any fermionic DM candidate that does not carry a conserved charge (or has an approximately broken conserved charge). These can be broadly classified into charged current and neutral current processes. 

A prospective Lagrangian for a neutral current process is given by~\cite{Dror:2020czw}
\begin{equation} 
{\cal L} \supset \frac{1}{ \Lambda ^2 } \chi \nu e ^c e +{\rm h.c.} ~ \Rightarrow ~ e ^- + \chi \rightarrow  e^- +\nu \,.
\end{equation} 
This process is well suited to detect DM with masses, $ m _\chi $, below the electron mass ($ m _e $) since the recoil energy deposited can be detectable by DM direct detection experiments even for $ m _\chi \sim {\rm keV} $ and exceed thresholds of some neutrino experiments for $ m _\chi $ of order hundreds of $ {\rm keV} $. A similar neutral current process can be present for nuclear scatterings~\cite{Dror:2019dib}. Furthermore, we note that to preserve lepton number, $ \chi $ must carry lepton number of $ + 1 $. Finally, a critical feature of fermionic absorption is DM is always unstable. It can decay through loop diagrams to combinations of neutrinos, photons, and, when its kinematically allowed, to heavier SM particles. This makes fermionic absorption primarily detectable for lighter DM candidates.

A prospective Lagrangian for a charged current process is given by,~\cite{Dror:2019dib}
\begin{equation} 
{\cal L} \supset \frac{1}{ \Lambda ^2 } \chi e   u d ^c +{\rm h.c.} ~ \Rightarrow ~ {} _{ Z} ^{ A} N + \chi \rightarrow {} _{ Z \pm 1} ^A N + e^b{\pm}   
\end{equation} 
where $ {} _{ Z} ^A N $ denotes a nucleus with atomic number, $  Z $, and mass number, $ A $. Charged current processes are analogous to induced $ \beta ^\pm $ decays traditionally considered for neutrinos but are arising from slow-moving DM. These explosive events convert an otherwise-stable nucleus to a nucleus with a higher or lower atomic number in the excited state. This nucleus will subsequently decay into its ground state along with  $ \gamma $. Just as for neutral current processes, the charge current process inevitably implies $ \chi $ decays however, these can be quite small, allowing for the possibility of detecting $ \chi $ with masses up to the pion mass.

\subsection{Other Cosmogenic Signals \label{sec:others}}

\subsubsection{Acceleration in astrophysical environments}

DM constituents carrying electric charge can be accelerated by astrophysical source~\cite{Chuzhoy:2008zy,Hu:2016xas,Dunsky:2018mqs,Li:2020wyl}, such as SN remnants. This can be naturally realized in the context of millicharged DM, which can appear in models with kinetic mixing~\cite{Holdom:1985ag,Foot:1991kb}.
The particles can be accelerated via first order Fermi acceleration mechanism (diffusive-shock acceleration) in analogy to conventional cosmic rays~\cite{Blandford:1978ky}. 
Fermi acceleration can accelerate ``dark cosmic rays''~\cite{Hu:2016xas} with electric
charge $\varepsilon e$, where $\varepsilon \leq 1$, to the maximum energy~\cite{Lagage:1983zz,Hillas:1984ijl} of $E_{\rm max} \sim \varepsilon e B U L$. For characteristic parameters of SN remnants, with the shock at
the end of the free expansion corresponding to acceleration length of $L \sim 3$~pc, magnetic field $B \sim 0.5$~mG and shock wave speed of $U \sim 0.1$, protons can be accelerated up to PeV levels while for millicharged particles this is lower by a factor of $\varepsilon$.
A distinct feature of dark cosmic rays is that their spectra, which depends on rigidity $R = p/Q$ of the particle with momentum $p$ and charge $Q$, is expected to follow a power-law~\cite{Hu:2016xas} - characteristic of cosmic rays accelerated in astrophysical environments. 
The resulting signatures can then be readily searched for in terrestrial laboratories~\cite{Hu:2016xas,Dunsky:2018mqs}, such as large neutrino experiments like Super-K as well as direct DM detection experiments.

\subsubsection{Atmospheric collider}

Cosmic rays bombarding the atmosphere results in copious production of particles, analogous to terrestrial colliders. The ``atmospheric collider'' has historically played a central role in neutrino physics, leading to discovery of neutrino oscillations~\cite{Super-Kamiokande:1998kpq}. Atmospheric collider is also a unique laboratory for exploration of BSM physics, including millicharged particles~\cite{Plestid:2020kdm,ArguellesDelgado:2021lek}, monopoles~\cite{Iguro:2021xsu}, light DM~\cite{Alvey:2019zaa, Su:2020zny} and heavy neutral leptons~\cite{Coloma:2019htx}. The resulting flux of boosted particles can then be readily searched for in terrestrial experiments, such as large DM and neutrino experiments. The atmospheric collider flux beam is always ON, it is robust and independent of local particle abundance, it extends over a broad spectrum from GeV to over PeV, and it is available for all terrestrial experiments. These features highlight atmospheric collider as an excellent laboratory for further exploration of new physics.  

\section{Neutrino Detectors}
\label{sec:exp}

A neutrino detector is designed to detect typically charged particles produced from neutrino interactions by collecting photons or charge proportional to the ionization energy loss, or photons from Cherenkov light.
Some detectors features on tagging neutrons.
In general, neutrino detectors allow precise measurements of physics quantities, such as event timing, deposited energy, event location, and direction of the particles.
In this section, we lay out several detection technologies, in examples of the corresponding experiments, used in neutrino detection and discuss the advantage and features of each type.

\subsection{Scintillation Detectors}

Neutrino detectors often use scintillator to detect the charged particles resulting from a neutrino interaction in the detector via their scintillation light.  Some detectors are segmented (MACRO, LVD, MINOS, and NO$\nu$A) for tracking and localization of background.  Others use continuous large volumes of scintillator, allowing the comparison of Cherenkov and Scintillation light to aid in particle ID and tracking. The downside is that light in a large volume will hit many photodetectors, making backgrounds harder to deal with.  Some detectors greatly purify their scintillator to limit low energy radioactive backgrounds.  KamLAND, Borexino, SNO+, and JUNO also go deep underground to limit cosmogenic backgrounds.  JUNO then dopes its scintillator with gadolinium: gadolinium has a high neutron capture cross-section, allowing the tagging of inverse beta decay neutrino interactions by looking for the coincidence between the positron and neutron capture.  The same technique allows large volume surface detectors such as Double Chooz, RENO, and Daya Bay to tag reactor neutrinos in spite of the cosmogenic background.  MiniBooNE gated on beam spills to reduce background by timing.

While all these detectors aim at various aspects of neutrino physics, they also have sensitivity to DM and other exotic particles. The simplest and so far most well-explored means are looking for neutrinos resulting from the annihilation of DM trapped in gravity wells (Section~\ref{sec:DMnus}).  The same signatures in the detectors could be used looking for BDM: any excess (or lack thereof) over ``ordinary''  neutrino signals~\cite{Giudice:2017zke,Dent:2019krz} could be interpreted in light of the models discussed in Section~\ref{sec:bdm}.  At higher energies, the larger water detectors in the next sections are more sensitive to this channel.  However, large low-threshold scintillator experiments will have mass larger than ``conventional'' DM experiments~\cite{Bramante:2018tos} with lower thresholds and better energy resolution than large experiments based on liquid-argon time-projection chambers (LArTPCs), such as DUNE, playing a complementary role to those detectors.

Note that the neutrinos that these detectors are designed to observe are also the observation channel for other BSM processes such as sterile neutrinos~\cite{MINOS:2020iqj}, neutrino magnetic moment~\cite{LSND:2001akn}, and Lorentz violation~\cite{MINOS:2012ozn}. If a core-collapse SN explodes somewhere in the Milky Way, the neutrino observations by all the experiments in Section~\ref{sec:exp} will test many BSM hypotheses both in particle and nuclear physics~\cite{Mirizzi:2015eza}: but that happens on average once or twice a century.  Light DM produced in an accelerator-based experiment's target also probes the dark sector~\cite{deNiverville:2011it,Batell:2014yra,deNiverville:2016rqh,DeRomeri:2019kic,Dutta:2019nbn,Buonocore:2019esg,Dutta:2020vop,Dutta:2021cip,Breitbach:2021gvv,Bhattarai:2022mue}.  If the nuclear suppression factor in carbon is more favorable than in oxygen, scintillator experiments could also be competitive with large water detectors in searches for baryon number violations via neutron-anti-neutron oscillations: the signature of such an event would be similar to the ``destructive'' DM interactions discussed in Section~\ref{sec:explosive}.

One exotic particle which scintillator detectors are well-suited to search for are magnetic monopoles.  These relics of the early universe are likely very massive and thus slow-moving.  Large water detectors are sensitive to the ionization produced by $\beta>0.9$ monopoles, but the more commonly hypothesized slower monopoles need scintillator light to detect. A monopole is likely to produce sufficient scintillator light for velocities down to $\beta > 10^{-4}$ to be identified via timing, and experiments tuned to detect a chain of single photo-electrons are sensitive to an order of magnitude lower speed.  While being deep underground helps reduce background and makes for a more efficient search~\cite{MACRO:2002jdv}, surface detectors add the ability to detect lower mass monopoles that will not have ranged out by the time they reach the detector~\cite{NOvA:2020qpg}.

\subsection{Water Cherenkov Detectors}

Water Cherenkov detectors (WCDs) are comprised of a central volume of clear water viewed by an array of photomultiplier tubes. Such a detector is ideal for searching for proton decay, the original motivation for IMB and Kamiokande, but also performs well in detecting neutrino interactions across the broad range of energies from MeV to TeV. There is only one such detector in the current generation: Super-K, which started operation in 1996 and continues today. The next generation detector, Hyper-K, is under construction and is expected to begin operation in 2027.

Neutrinos are detected by their interaction in the central volume. Cherenkov rings are reconstructed for the charged particles that are above Cherenkov threshold. The interaction vertex is localized to around 30 cm for 1 GeV scale events. At energies from 5 MeV to 100 MeV, neutrinos that elastically scatter off atomic electrons are detected, but restrictions such as time coincidence (such as a SN explosion) or the angle from a point source (such as the Sun) must be employed to get a measurement above background due to radioactivity. At energies from 100 MeV to 1 GeV, the event patterns are clear and free from radioactive background, but the neutrino direction is poorly known due to the large mass of the recoil nucleon. At higher energies, of order 10 GeV, the event patterns are less distinct, high energy muons may exit the detector, but the neutrino direction is better reconstructed. High energy (100 GeV) muon neutrinos that interact in the rock around the detector also provide a substantial sample, with good astrophysical pointing resolution of a few degrees, but these events are limited to charged current muon neutrino interactions. The basic assumption is that the neutrinos detected by Super-K or Hyper-K are due to atmospheric neutrino production, but within this sample one may look for neutrinos of cosmogenic origin.

For ``standard'' indirect DM detection via neutrinos, Super-K uses a complete model of atmospheric neutrino interactions, including the effect of neutrino oscillation, to look for an excess of events pointing towards gravitational wells such as the GC~\cite{Super-Kamiokande:2020sgt}, Sun~\cite{Super-Kamiokande:2015xms}, or center of the Earth~\cite{Frankiewicz:2015zma}. Thanks to a large target mass of 22.5 kton, long exposures of up to 15 live-years, and sensitivity to charged-current neutrino interactions of all flavors below 10 GeV, Super-K has set stringent limits on annihilation of DM for masses below 100 GeV, with the GC example shown in \autoref{fig:nu_spectra}. Marginal improvements will be possible with the final Super-K data set. Hyper-K can resume the study with 186 ktons of mass.

Due to the Cherenkov threshold, ``standard'' WIMP-nucleon scattering is not observable. However, BDM affords an opportunity to find energetic recoiling particles. As with the indirect DM searches described above, atmospheric neutrinos comprise a considerable background, one that can be battled by pointing at gravitational wells. For BDM that couples mainly to quarks, the signature is an excess of neutral-current like interactions that point back to a possible source. The interactions may be single protons above Cherenkov threshold, or energetic neutral-current events that must be separated from charged-current events. This search is in progress. For BDM that couples strongly to electrons, the signature of a purely electromagnetic shower affords considerable background reduction. The Super-K search vetoes on the presence of the $\pi-\mu-e$ decay chain and any sign of neutron production and capture. This search was negative in 7-years of SK-IV data~\cite{Super-Kamiokande:2017dch}.

Water Cherenkov experiments have also searched for exotic cosmic particles such as Q-balls~\cite{Super-Kamiokande:2006sdq}, fractionally charged particles~\cite{Kamiokande-II:1990hos}, and magnetic monopoles~\cite{Becker-Szendy:1994kqw,Super-Kamiokande:2012tld}. These exotic signatures provide opportunities for creative analyses that broaden the scientific research program and one of which might just come to fruition someday.

\subsection{Long-String Water Cherenkov Detectors}



Gigaton volume neutrino detectors can be constructed by integrating long strings of optical sensor modules with precision timing into naturally occurring media with good optical properties, such as deep ocean water, freshwater, or glacial ice. Neutrinos are detected via the emitted Cherenkov radiation following neutrino interactions. Recent scientific breakthroughs, such as the discovery of high-energy astrophysical neutrinos~\cite{Aartsen:2014gkd,Aartsen:2013jdh} and the recent observation of neutrino emissions in the direction of blazar TXS~0506+056~\cite{IceCube:2018dnn,IceCube:2018cha} by the IceCube Neutrino Observatory, have established neutrino telescopes as discovery machines.

The simple detector designs of neutrino telescopes, consisting of 3-dimensional arrays of sensor modules make them multi-purpose detectors with very diverse science programs~\cite{Rott:2021sgf}. This concept has been successfully employed by IceCube at the geographic South Pole, the ANTARES neutrino telescope~\cite{ANTARES:2011nsa} in the Mediterranean sea, and the Lake Baikal experiment and its successor the Baikal GVD telescopes~\cite{Avrorin:2018ijk}. IceCube can detect neutrinos from 10~GeV to beyond PeV energies, as well as galactic SN burst neutrinos at 10~MeV in the format of thousands of 10-MeV neutrino interactions occurring within the time scale of a second. The IceCube Upgrade, currently under construction, and prospects for creating an IceCube-Gen2 facility~\cite{Aartsen:2020fgd}, will expand the energy coverage and the scientific scope even further. International efforts are underway to construct new telescopes in Europe (KM3NeT-ORCA near Marseille and KM3NeT-ARCA near Sicily)~\cite{Adrian-Martinez:2016fdl}, in Siberia (Baikal-GVD)~\cite{Avrorin:2018ijk}, and in Antarctica (IceCube-Gen2)~\cite{Aartsen:2020fgd,Aartsen:2014njl}; a test site is being explored at the Ocean Networks Canada in the Pacific Ocean (P-ONE)~\cite{Agostini:2020aar}.

BSM rare process searches in these detectors are especially powerful, thanks to their strong statistical precision allowed by large volumes. 
Large volume allows them a unique program to study BSM phenomena at an energy scale beyond the reach of LHC, and reaching the highest energy amongst terrestrial detectors.
Some of the disadvantages of these detectors are the sparse photo-coverage, uncertain optical properties in the natural medium, and the limited energy reconstruction power of the high-energy muon-neutrino sample.

Nevertheless, very stringent bounds on heavy decaying dark matter have been achieved with IceCube excluding lifetimes of up to $10^{28}$~seconds depending on the decay mode~\cite{Aartsen:2018mxl,Arguelles:2019boy,Rott:2014kfa}. Searches for self-annihilating DM exploiting a variety of targets with large DM accumulations, such as the galactic halo~\cite{Abbasi:2011eq,Yuksel:2007ac}, GC~\cite{Aartsen:2017ulx,ANTARES:2020leh}, or clusters of galaxies~\cite{Aartsen:2013dxa,IceCube:2021sog}, have been conducted as well. DM searches from the center of the Sun~\cite{Gould:1987ir,Gould:1991hx,Wikstrom:2009kw,Rott:2011fh,Danninger:2014xza,Liu:2020ckq,Bauer:2020jay}, have resulted in very competitive bounds~\cite{Aartsen:2016zhm,ANTARES:2016xuh} with minimal dependence on the underlying DM velocity distributions~\cite{Danninger:2014xza,Choi:2013eda,Nunez-Castineyra:2019odi}. The detailed timing and energy spectrum of neutrinos from a galactic core-collapse SN can be used to probe sub-GeV dark sector or axion-like particles~\cite{Chang:2018rso,Fischer:2016cyd,Benzvi:2017ncw}.
Searches for BDM can probe a variety of scenarios with DM masses ranging from PeV scales to sub-GeV particles~\cite{Bhattacharya:2014yha,Bhattacharya:2016tma,Kopp:2015bfa,Super-Kamiokande:2017dch,Kim:2020ipj}.



\subsection{Liquid-Argon Time-Projection Chambers}
\label{sec:LArTPC}

Large detectors based on Liquid-Argon Time-Projection Chambers, 
championed as neutrino detectors by the ICARUS experiment~\cite{ICARUSLNGS} at LNGS, Italy,
have been the technology of choice for several experiments based on neutrino beams in the U.S.A.,
namely the three ones in the SBN program at Fermilab~\cite{SBN,MicroBooNE,SBND,ICARUS}
and DUNE~\cite{DUNE}, where both the far detector~\cite{DUNEFD} and a component of the near detector~\cite{DUNELArND} feature it.

The most apparent strength of LArTPC is the imaging of interactions with a spatial resolution to the millimetre,
which enables detailed reconstruction of physics events with high particle multiplicity,
and achieves unparalleled resolution on the interaction vertex.
The detection mechanism consists of allowing the track of electrons ionized by charged particles crossing argon
to drift to an instrumented anode. This process typically takes a time on the order of a millisecond
(with drift velocity of $\approx 0.6\,\textnormal{ms/m}$).
LArTPCs can image all charged particles
and can detect neutrons when their secondary interactions produce ionization,
and photons as they produce a $e^{+}e^{-}$ pair or kick out electrons.
The ability to track protons is an improvement over water detectors;
on the other hand, the latter have lower threshold for detecting electrons.
LArTPCs cannot distinguish the sign of charge unless magnetized,
a challenging solution that has not been deployed on large LArTPCs so far.

The mode for operating a LArTPC greatly differs for surface and underground detectors.
On surface, as it is the case for the SBN program detectors and the DUNE prototypes~\cite{ProtoDUNE} hosted at CERN,
signals from cosmic rays dominate even when using the typical meter-deep concrete shielding.
This constant flux frustrates a trigger exclusively based on the activity observed in the detector
and forces an acquisition synchronized with the pulsed neutrino beam, with a duty cycle of $\approx1\%$ deliberately undercutting the detection of cosmogenic activity.
Deep underground environment instead allows to trigger the acquisition according to the signal observed in the detector.
In this environment, only atmospheric neutrinos and detector radioactivity constantly leave signals. Both DUNE near and far detector fall in this category.



The reconstruction of incoming particle direction, essential to point back to its source, is achieved as the combination of momenta of all the observed particles.
The information-rich imaging of LArTPC enables detailed reconstruction of the interaction, resolving the single particles.
The advantages of this technology include the precise energy reconstruction from stopping range when particles are contained in the detector,
the low detection threshold for protons (of the order of 50\,MeV),
and the high resolution of the location of the interaction (better than one centimeter).
This becomes an essential tool when searching for a point-like cosmic source, even more so when specifically targeting a known one (e.g.~the GC).
The detectors on neutrino beams have been designed to maximize the containment of interactions from their beam, with a shape elongated in the beam direction, approximately horizontal.
That choice penalizes the containment of cosmogenic activity, which has either no preferential direction or, if shielded by the Earth, is predominantly downward,
impacting the isotropy of the sensitivity and the analyses based on contrast of upward-/downward-going and night/day interactions (Section \ref{sssec:OnOffMethods}).



The current state of the art of LArTPC technology, DUNE far detector, includes 70~kton of liquid argon.
While increasing the size of a LArTPC is a major challenge,
the modular solution, already planned in DUNE,
resolves the technical challenges at the price of reducing the fraction of well-reconstructible active volume. 




\section{Analysis Strategies}
\label{sec:strategy}

\subsection{Signal Simulation}
\label{sec:sim}

In the search for the scattered electron or decay channels,
simulation of the cosmogenic DM and exotic particles can be as simple as simulating the electron scattered by those particles, and the visible decay products.
However, in the case of nucleon scattering, and of anisotropic detectors such as the DUNE far detector, more sophisticated simulation is required for the nucleon interactions and propagation and for the containment of the final state particles. 

The DM and exotic particles scattering off of nucleons are the most challenging interactions to simulate.
As the detection range of neutrino detectors are above the MeV regime -- most of the far detectors for accelerator-produced neutrinos are optimized to the GeV scale, -- energy transfers are around the QCD scale, a complicated energy scale to deal with.
In addition, hadrons produced in these ``core'' interactions can scatter as they exit the nuclear debris, a phenomenon also known as Final State Interactions (FSI). This scattering significantly distorts the particles produced from the core interactions and their kinematics.
Currently, for a broad class of DM models, namely those with a spin-1 mediator, a tool has been developed to simulate hadronic scattering \cite{Berger:2018urf}. 
This tool is built within the \texttt{GENIE} software \cite{Andreopoulos:2009rq}, making it straightforward to test different nuclear models and to connect with detector simulation pipelines.

Underground, massive neutrino detectors, which are typically not immersed in a magnetic field, the containment of final state particles has a relevant impact on particle identification and energy resolution.
While the detector geometry and containment of the final state particles are well-established in the official analysis framework pipeline of all the experiments, it is important to have a mechanism to easily, seamlessly integrate these effects with the simulation of the cosmogenic DM and exotic particles, lowering the barrier to initiate search for such BSM physics.

While more details about simulation tools are discussed in the Snowmass 2021 White Paper~\cite{Batell:2022temp}, we reiterate the relevance of the simulation for the upcoming generation of cosmogenic BSM searches.

\subsection{Triggering}
The neutrino detectors designed for measuring solar, reactor, or atmospheric neutrinos use a light-based trigger, either from scintillation light or Cherenkov radiation emitted by the particles resulting from the neutrino interaction. Therefore, these detectors are expected to trigger also on cosmogenic DM signals as well, provided that the energy deposited in the detector is at least comparable to the neutrino signal.

Accelerator-based neutrino experiments, especially those which are not big enough to include a dedicated physics program with atmospheric neutrinos or cosmic rays (e.g.~MicroBooNE, SBND, ICARUS), rely on the coincidence with the accelerator signals for triggering, which prevents them from acquiring cosmogenic DM signals unless they are contained in unbiased triggers. 

\subsection{Reconstruction}
The technical procedure of reconstructing signal and background in a detector highly depends on the design and technology on which that detector is based.  However, most reconstruction procedures can be broadly separated into a few distinct steps:
\begin{itemize}
    \item \textbf{Signal processing:} The initial preparation of measurements on a channel in a readout window.  This typically involves removing detector effects and noise.
    \item \textbf{Hit reconstruction:} Locating and characterizing signals on a channel where the measured signal goes above some threshold.
    \item \textbf{Clustering:} Grouping together reconstructed hits, based on some criteria (usually proximity-based).  The clusters are typically early representations of particles.  Depending on the detector technology, the clusters could be 2D or 3D in nature.
    \item \textbf{Characterization:} Application of more advanced procedures to reconstruct properties of the particles, e.g.~application of a Kalman filter, to a cluster to reconstruct a particle's trajectory.
\end{itemize}
In neutrino experiments that also includes DM or exotic particle searches as a goal, it is usually possible to reuse and adapt the neutrino reconstruction, rather than design targeted reconstruction from the ground up, for such additional searches.  An example of this is in LArTPCs where the Pandora pattern recognition suite~\cite{MicroBooNE:2017xvs} was designed to reconstruct neutrino interactions, but is also used for DM and exotic particle signals.  Important but typically small amendments usually need to be made when adapting the reconstruction, such as ensuring the vertex reconstruction can account for the isotropic nature of DM and exotic signals, in contrast to the typically forward going nature of beam-induced neutrino interactions.

\subsection{Analysis Strategies}

\subsubsection{Overall strategy}

Cosmogenic DM searches typically seek out a steady signal having been formed throughout the long history of the Universe.
A principal strategy in searching for such signal is to exploit directional information, to separate signal flux from backgrounds and to localize the source. Matching cosmogenic DM particles or neutrino yields from them with their sources is relatively straightforward compared to charged cosmic ray's case, as they are not expected to experience the electromagnetic force on the way.

In the scenarios of DM annihilation or decay, the resulting signal spectrum depends on particle-physics properties, such as the DM mass and the branching ratios to visible and invisible channels. 
By reconstructing the energy spectrum of the detected particles, one can deduce the energy spectrum of the parent neutrinos or BDM, therefore the properties of the DM.
In case of direct annihilation to a pair of neutrinos ($\chi\bar{\chi} \rightarrow \nu\nu$) or annihilation channels inducing stopped meson decay in the Sun, \cite{Rott:2012qb,Bernal:2012qh,Rott:2015nma} 
the produced neutrinos will have a mono-line spectrum resulting in supreme separation power to background events \cite{Lindner:2010rr}. 
With detectors and strategies which can resolve a line signal or a sharp spectral edge, it is plausible to deduce the mass of the DM from the spectrum produced by neutrino or hard channels, the latter of which results in a neutrino spectrum dominated by neutrinos carrying a large fraction of the available energy from the decay or annihilation.

Recorded events in neutrino experiments 
are dominated by cosmic-ray muons generated by cosmic-ray nuclei interacting in the atmosphere. To reject them, muons coming below the horizon (``upgoing tracks'') are accepted as the Earth filters the cosmic-ray muons.
If the vertex point is contained in the detector (``starting track''), down-going particles are also accepted as a neutrino or dark cosmic ray candidate. 
Remaining particles spanning a wide range of energies are atmospheric neutrinos produced by cosmic-ray interactions as well, being the major and unavoidable background for exotic particle searches.
Super-K has started a new phase of its operation in July 2020 with gadolinium-doped water \cite{Mori:2013wua}. The enhanced neutron capture in the new running phase will reduce contamination from neutrino charged-current interaction in the searches of recoiled electrons by neutral particles or millicharge particles.
Solar neutrinos and radioactive backgrounds come in for low-energy searches, requiring a careful treatment especially in case of marginal directionality.
Detectors with limited directional sensitivity may use other tactics such as timing information to reduce background.
Unique topologies made by exotic particle interactions are also useful for background elimination, especially for LArTPCs, owing to its millimeter spatial resolution.
Further, new reconstruction and background rejection techniques are needed to improve the sensitivity and the uncertainties on the remaining backgrounds need to be studied.



\subsubsection{On-Off methods}
\label{sssec:OnOffMethods}


While a signal comes from the direction of the source, background like atmospheric neutrinos reach the detector from all directions and can be significantly reduced by accepting only the events for which the reconstructed incoming particle's direction points to the source position at the time of the event.
ON and OFF-source analysis is a data-driven approach which has least model-dependency as it 
allows the analysis to be independent of the background Monte Carlo (MC) simulations and related systematic uncertainties as they should equally affect ON and OFF-source regions. 

DM density distribution in a cosmogenic source affects angular distribution of the signal. Since DM decay only depends linearly on the DM density where annihilation depends quadratically, its signal exhibits a broader angular distribution.
In case of a spread signal, after normalizing the number of events to the angular size, the difference in the events between On- and Off-source regions:

\begin{equation}
    \Delta N = (N^{BG}_{On} + N^{sig}_{On}) - (N^{BG}_{Off} + N^{sig}_{Off}) \sim N^{sig}_{On} - N^{sig}_{Off} \sim \Delta N^{sig},
\end{equation}

takes account the fact that the Off-source region contains nonzero fraction of DM component.


To keep majority of the signal events after applying a directional cut, good angular resolution is required.
The angular resolution depends on several factors, including kinematics leading to a spread in direction of outgoing particle that is detectable, photo-coverage, and reconstruction algorithm.
Since the angular correlation between the incoming BDM or neutrino and the outgoing particle depends on the energy of BDM or neutrino, the region of interest is often determined by the properties of DM, such as mass and branching ratio. 
In the DM-induced neutrino searches, track-like signal left by high-energy muons are widely used for its good directionality. At high energy above 10 TeV, muon neutrinos interacting with hadrons via charged-current quasi-elastic scattering emits muons in the forward direction within $<$ 0.1$^\circ$. The effective target volume increases as neutrino energy increases, because both the cross section and the distance traveled by the muon before entirely losing its energy are proportional to the neutrino energy, stretching the target volume far below the detector.
At lowest neutrino energies below 15 MeV, positrons are emitted roughly isotropically \cite{Vogel:1999zy}. However, hadronic recoil is largely forward peaked, therefore detectors with capability of determining proton's direction, such as DUNE, can potentially utilize the information for better pointing ability \cite{Rott:2016mzs,Rott:2019stu}. 
Higher-energy neutrino or BDM also accompany lower background noise level as the atmospheric neutrino flux drops proportional to $E^{-2.7}$.  


Limits set on the excess of events in the On-source region can be interpreted as limits on the DM scattering cross section, annihilation cross section or lifetime depending on the tested model.
In a quasi-model-independent approach, analysis is done for various assumed DM masses and for the softest and the hardest annihilation or decay spectrum (or group of them), i.e. $b\bar{b}$ (and $W^+W^-$) and $\nu\bar{\nu}$ (also $\mu\bar{\mu}, \tau\bar{\tau}$) channel with 100\% branching ratio \cite{Super-Kamiokande:2011wjy}.
Realistic scenarios are expected to fall within the extreme results and the effect of testing several models on the signal significance is taken into account in terms of a trial factor.


In cases with limited directional information as for the scintillation detectors, modulation of the signal strength can be useful. 
Strongly Interacting Dark Matter with hadrons could be searched for with cosmic-ray boosted signal flux \cite{Cappiello:2019qsw}. 
PROSPECT, located on Earth's surface, has probed large DM-nucleon cross sections by searching for significant degree of diurnal sidereal modulation, where background electrons were efficiently removed by strong pulse shape discrimination of the scintillation detector \cite{PROSPECT:2021awi}.
Point-source searches without an assumption on the source position in the sky, often called ``unbiased searches'', can be useful for searches of unknown DM source.

\subsubsection{Signal+background fitting}

Log-likelihood ratio test is a widely used technique to distinguish between two mutually exclusive hypotheses, having been applied in broad categories of atmospheric neutrino data analyses.
To evaluate a hypothesis of the presence of DM-induced events in the data, the likelihood ratio, or almost equivalently, the $\Delta\chi^2$, is selected as a test statistic:

\begin{equation}
    \Delta\chi^2 = \chi^2(\textnormal{null hypothesis}) - \chi^2(\textnormal{best fit}) = \chi^2 (\beta=0) - \chi^2 (\beta_{min}),
\end{equation}
where $\beta$ represents the strength of the signal and $\beta_{min}$ indicates the parameter value at the $\chi^2$ minimum for the best-fit point. 
$\chi^2$ is defined as: 
\begin{eqnarray}
\chi^2 = \sum_{i=1}^{n}\chi_i^2 = \sum_{i=1}^{n}2 (N_i^{exp} - N_i^{obs} + N_i^{obs} \log \frac{N_i^{obs}}{N_i^{exp}})
\end{eqnarray}
where $i$ is the bin index and $\chi_i^2$ is the chi-square of the $i$-th bin and $
N_i^{exp} = N_i^{BG} + \beta N_i^{signal}$ is the expected number of events in the $i$-th bin; the bin index $i$ can be multi-dimensional, spanning several observables.

Candidate events 
are distributed in the preferred angular distribution defined as solid angle from a cosmogenic source.
These events are compared to the combination of the signal and background MC simulation events, where signal MC is produced by running a simulation of DM production and its propagation through dense medium and vacuum, and the interaction inside the detector. The signal is simulated for different DM particle-physics properties, rendering the fit to run multiple times for the total number of different DM hypotheses tested.

For lower-energy signal induced by lighter DM, since they have diminished angular resolution, finding an appropriate angular cut with a good signal-to-background ratio as keeping a high signal efficiency is challenging, in particular with the increased flux of atmospheric neutrinos at lower energy region \cite{Feng:2008qn,Kappl:2011kz,Rott:2011fh}.
In the multi-bin log-likelihood approach, the signal selection cuts can be relaxed compared to a cut-and-count analysis, resulting in increased signal statistics.
In addition to the distinctive angular distribution of the signal, difference between the energy spectra of the DM-induced signal and the atmospheric neutrinos can be utilized, which is especially useful in searches with low-energy contained events.
Accepting signals in all-flavor event sample will not only increase statistics, but also adds discrimination power against background atmospheric neutrinos, as well as potential information of dark matter branching ratio \cite{Kumar:2009ws}.

Another advantage of the Likelihood Ratio approach is the possibility to treat nuisance parameters in a straightforward manner.
In the ``pulled technique", the nuisance parameters representing detector and other systematic uncertainties are fitted together with the strength of the signal, 
with the $\chi^2$ defined as:

\begin{eqnarray}
\chi^2 =  \sum_{i=1}^{n} \chi_i^2 + \sum_{j=1}^m \frac{\epsilon_j^2}{\sigma_j^2}
\end{eqnarray}

where $\sigma$ is the $j$-th systematic uncertainty and $\epsilon_j$ is the ``pull" to it and the $N_i^{exp}$ is defined as:

\begin{eqnarray}
N_i^{exp} = (1 + \sum_{j=1}^k f_j^i \epsilon_j) N_i^{BG} + \beta(1 + \sum_{j=k+1}^m g_j^i \epsilon_j) N_i^{signal}
\end{eqnarray}

where $f_j^i$ and $g_j^i$ are the response coefficients of the $i$-th bin to the $j$-th systematic uncertainty for expected background and expected signal events, respectively.
The pull approach is especially practical to deal with large number of systematic uncertainties in the atmospheric neutrino production, interaction and reconstruction.




\subsubsection{Background modeling and uncertainty}

Data-driven approaches are free from background modeling.
Otherwise, it is important to reduce background-related uncertainties by improving background modeling accuracy.
The atmospheric neutrino flux is predicted by 3D simulation of air showers with inputs such as air density models, geomagnetic models, and primary particle fluxes based on cosmic-ray satellite data \cite{Honda:2006qj,Fedynitch:2018cbl}.
The uncertainty of the atmospheric neutrino flux calculation is about 7\% from 1 to 10 GeV, which increases to $\sim$25\% at 1 TeV \cite{Honda:2006qj}.
At high energies, hadronic interactions inside the air showers are only accessible through simulation, becoming a major source of the uncertainty of the flux calculation.
In order to improve the accuracy of the hadron interaction model, 
efforts to incorporate accelerator hadron measurements into atmospheric neutrino simulation are on-going \cite{Sato:2020hrb}.

In addition, next-decade experiments will face a background contribution from neutrinos produced in the atmosphere of the Sun by cosmic-ray collisions, also known as ``solar atmospheric neutrinos,'' resulting in a solar atmospheric neutrino floor~\cite{Arguelles:2017eao,Ng:2017aur,Edsjo:2017kjk,Masip:2017gvw}. Attempts to measure the solar atmospheric neutrino flux are on-going \cite{In:2017qma,Aartsen_2021,IceCube:2021koo,ANTARES:2022azv}.



\section{Complementarity of Different Experiments}
\label{sec:complement}

The emerging theory models, as the ones discussed in Sections.~\ref{Sec:models} and \ref{sec:modeloverview}, can be tested at detectors based on different technologies which complement one another; on the other hand, if a signal or an anomaly is observed, a confirmation delivered by a different detector --- hence, with different systematic uncertainties --- is necessary.

Underground, multi-kiloton-scale or larger neutrino detectors have shown great sensitivity to the signals from cosmogenic DM and exotic particles, with the advantage of detecting small signal fluxes, and the future detectors have potential to improve the sensitivity and probe wider, more sophisticated parameter space, wherein the answer to BSM physics may sit. 
In addition, the complementary information measured from different detectors will contribute to multi-messenger astronomy.
It is desired to utilize existing and planned detectors, irrespective of their primary physics goals.


\subsection{Neutrino Detectors based on Different Technologies}
\label{sec:comple_nuDet}

As discussed in Section~\ref{sec:exp}, neutrino detectors are sensitive to different energy ranges and different physics quantities depending of their technology of choice.
The parameter space probed by those neutrino experiments is therefore complementary, and it is desirable to collect all the information that we can obtain from them.

The DUNE far detector based on the LArTPC technology, leveraging on their good reconstruction of hadron direction~\cite{Berger:2019ttc} and relatively low detection threshold, may use hadrons produced from exotic or neutrino interactions as an effective channel for exotic searches, such as elastic and inelastic BDM searches. Scintillation detectors such as Borexino and JUNO also have high discovery potential.
In addition to reduced signal acceptance of hadronic signal due to high Cherenkov threshold of proton in the water ($\sim$1.2 GeV), WCDs like Super-K also suffer from challenges in the identification of recoiled protons \cite{Super-Kamiokande:2009kfy,Berger:2014sqa,Agashe:2014yua}.
WCD has lower threshold for detecting electrons, being an excellent tool to seek sub-GeV signals. Scintillators typically with lower energy threshold, may even access to sub-MeV signals. 

In the BDM models involving ``upscattering'' processes or bremsstrahlung of a dark gauge boson, a pair of lepton and anti-lepton is produced together with a recoiling target, allowing further reduction of the neutrino background \cite{Kim:2016zjx,deGouvea:2018cfv,Kim:2019had}. 
In general, for signals containing multiple visible particles in the final state, separation power among the outgoing particles is critical, requiring the use of a detector with excellent angular resolution and position resolution 
like DUNE. 
It is also shown that an energy-density-based analysis can help recognizing merged or overlapping tracks \cite{DeRoeck:2020ntj}.

Some of the millicharged particles would leave multiple energy depositions while traversing the detector, instead of a strong single-hit signal~\cite{Magill:2018tbb}.
Depending on the charge, these signals can be detected as faint tracks or as a sequence of electrons displaced by multiple scattering.
The time-correlated, aligned multi-hit search will benefit from significantly reduced background, allowing to probe smaller charges with low-energy threshold detectors such as JUNO. It is also demonstrated that detectors with longer path lengths such as IceCube are well-suited for multiple-scattering searches \cite{ArguellesDelgado:2021lek}.

\subsection{Complementarity to Direct Dark Matter Detectors}
\label{sec:comple_directDM}

The standard two-component BDM scenario (see Section~\ref{sec:twocomponentbdm}) is one of the actively investigated scenarios for which searches in neutrino experiments and in conventional DM experiments are complementary. 
For example, \cite{Kim:2020ipj} shows that different experiments, such as DUNE, Super-K/Hyper-K, IceCube, and DarkSide, are best suited for probing different regions of BDM parameter space, and provides guidelines to optimize search strategies. 
For a significant fraction of parameter space in the BDM models, the BDM flux is much smaller than the halo DM flux, and the BDM interactions in the detectors deposit energy above MeV scale.
Underground, multi-kiloton-scale neutrino detectors therefore have competitive sensitivity in searching for such BDM signals, and can be complementary to the halo DM search in the direct DM detection experiments.
On the other hand, in the particular scenario with sub-GeV halo DM, the BDM flux in the two-component DM models is generally anticipated to be significantly more intense, as suggested by \autoref{eq:fluxAnnihilation}, and the energy of the visible particles produced by the BDM interaction is at the keV regime\footnote{More precisely, these visible particles can also have energy up to a few MeV, where both neutrino detectors and direct DM detectors are sensitive to~\cite{Giudice:2017zke,Kim:2020ipj}.}~\cite{Cherry:2015oca,Giudice:2017zke,Kim:2020ipj}.
Unlike the other scenarios, the multi-ton-scale experiments for DM direct detection benefit from their keV detection threshold for such BDM signals, while the relatively small target mass can be offset by the large BDM flux.

Detection of BDM in DM direct detection experiments~\cite{Cherry:2015oca,Giudice:2017zke,Kim:2020ipj} also received particular attention in the context of the XENON1T low-energy excess in the electron recoil channel~\cite{XENON:2020rca}, as it is less compatible to the signals that models of conventional halo DM interacting with electrons would predict. 
In parallel, a class of DM interpretations assumes the existence of fast-moving or boosted DM consisting of a subdominant fraction of the DM energy budget~\cite{Kannike:2020agf,Wang:2019jtk,Fornal:2020npv,Cao:2020bwd,Jho:2020sku,Ko:2020gdg,Alhazmi:2020fju,DelleRose:2020pbh,Cao:2020oxq,Jia:2020omh,Borah:2021yek}, as also alluded in Section~\ref{sec:BDMsearches}.
As more ton-scale DM direct detection experiments have just started operating, it is expected that they can explore certain regions of BDM parameter space to which the neutrino experiments are less sensitive~\cite{Kim:2020ipj}. 

In a similar manner, ultra high-energy BDM signals ranging from PeV to EeV can be observed in ultra-high-energy cosmic-ray detection experiments.  For example, it was demonstrated that the ANITA anomaly~\cite{ANITA:2016vrp,ANITA:2018sgj} can be explained by the BDM in the decaying two-component DM scenario~\cite{Heurtier:2019rkz} (see Section~\ref{sec:twocomponentbdm}).

\subsection{Complementarity to Searches for Signals Produced by Accelerators}
\label{sec:comple_accelerator}

An important avenue in probing the models and scenarios predicting cosmogenic signals lies in the searches in accelerator-based experiments, especially when the associated mass scale is within the reach of the accelerator-based experiments.
While cosmogenic signals come from nature, signals produced by accelerators are artificial and therefore the experiment environment and the background activities can be much better controlled.  
The two approaches with natural and artificial sources hence provide complementary information in probing BSM physics. 


Accelerator-based neutrino experiments typically have neutrino beams from meson decays, where the mesons are produced by protons colliding on a fixed target.
Those experiments, with the near detectors or short-baseline detectors a few hundred meters away from the target, have potential to detect BSM particles produced by meson decays in the flight (e.g.\ the near detectors of DUNE and T2K/T2HK, as well as SBND, MiniBooNE, MicroBooNE, and ICARUS), or by meson decays at rest (e.g. LSND, KARMEN, COHERENT, CCM and JSNS$^2$).
Together with the fixed target experiments, such as SeaQuest, NA62, NA64, and LDMX,
these experiments feature high-intensity particle beams and therefore even weakly interacting particles can be copiously produced. 
See, for example, NF White Papers~\cite{Arguelles:2019xgp,Anchordoqui:2021ghd,DUNE:2022aul,Toups:2022yxs,Toups:2022knq,Batell:2022temp} and RF White Papers~\cite{Anchordoqui:2021ghd,Toups:2022yxs,Toups:2022knq,CraigGroup:2022swb,Apyan:2022tsd} for Snowmass 2021 for more dedicated summaries of related activity.


On the other hand, collider experiments are better suited for producing GeV to TeV-scale particles with a sizable coupling to SM particles. 
Lepton collider experiments, such as BaBar, Belle-I/II, and LEP, have played an important role in probing BSM physics models, including the dark-sector scenarios, as summarized in an EF White Paper for Snowmass 2021~\cite{belle2}.
Dedicated detectors for the search for long-lived particles, such as CODEX-b, FASER/FASER$\nu$, MATHUSLA, MILLIQAN, and M\"{o}EDAL, can be 
sensitive to BSM physics models, as illustrated in the Snowmass-2021 White Papers~\cite{Anchordoqui:2021ghd,Alimena:2022hfr,Aielli:2022awh,MATHUSLA:2022sze}.
Moreover, hadron colliders, such as the LHC, offer a variety of opportunities: ATLAS, CMS, and LHCb have set competitive limits on dark-sector particles with the mass scale at the GeV-TeV regime for a wide range of channels, most notably with the
mono-X searches aiming at the final states containing a single jet, a photon, a W/Z gauge boson, or a Higgs boson plus momentum imbalance.
They have also put more endeavor in the search for long-lived particles.
Relevent Snowmass-2021 White Papers can be found in~\cite{Chekanov:2020xco,Franceschini:2022vck}.

As a number of experimental efforts and phenomenological developments have been made in search for dark-sector or exotic particles produced by accelerators,
it is desired to further investigate and develop the complementarity to searches for cosmogenic signals in underground, massive neutrino detectors.

\subsection{Outlook}

It is vital to have complementary search for cosmogenic BSM physics, and it will be imperative to have confirmation from similar and complementary experiments if evidence of BSM physics is observed.
Close collaboration and conversations between theory and experiment communities will definitely enhance the utilization of existing experiments and proposed facilities.
Further, it is desired to have more studies sketching the landscape of the existing results and the prospects of the future developments:
the concise interpretations of the existing results from complementary experiments will enhance our understanding of the status of the BSM search, while the studies based on the prospect facilities will inform their design.
In addition, it is important for existing data to be made accessible to an extent that allows comparisons and analyses of correlation.
Similar to the case of SN neutrino detection, where different detector technologies are sensitive to different flavors of SN neutrinos, the combination of those results will provide us with a more comprehensive picture, and will contribute to the multi-messenger astronomy.

\section{Conclusions}
\label{sec:conclusion}

This White Paper focused on the searches, in neutrino experiments, for the signals originating from the cosmogenic DM and exotic particles in the present universe. 
As summarized above, various new physics models and dark-sector scenarios predicting non-conventional cosmogenic signals have been proposed and developed in the past decade, (partly) motivated by the fact that no conclusive DM-like signals have been observed in DM detection experiments.
The signal flux expected in those models and scenarios is often small, and therefore, neutrino experiments equipped with kiloton-scale or larger detectors have great potential on such searches.
Several operating neutrino experiments have already published their results, setting new limits for the related DM models~\cite{Super-Kamiokande:2017dch, COSINE-100:2018ged, PandaX-II:2021kai, CDEX:2022fig}.
It is expected that the importance of cosmogenic DM and exotic particle searches will keep growing in the upcoming decades.

The recent development on detector technologies for neutrino experiments, in particular the LArTPC technology, offers searches in complementary parameter space.
As neutrino measurements are entering the precision era, optimizing the analysis strategies and tools for the cosmognic signal searches will enhance their sensitivity. 
This requires more dedicated, sophisticated event simulation, triggering, reconstruction algorithms, background characterization, and analysis techniques.
Based on the existing techniques summarized in this White Paper, a wider range of developments has just started, and we strongly encourage to dedicate more resources from now.

We emphasized the importance of the complementarity between neutrino, direct DM, and accelerator-based experiments in search for cosmogenic DM and exotic particles.
It is crucial to sketch a landscape of existing results and of prospect sensitivities;
further, an established protocol of combining data from multiple experiments will shed light on the unknowns and empower the multi-messenger astronomy.

\section*{Acknowledgments}
AD was supported by
the U.S. Department of Energy under contract number DE-AC02-76SF00515.


\renewcommand{\refname}{References}


\bibliographystyle{utphys}

\bibliography{tex/ref}

\end{document}